\documentclass[fleqn,usenatbib]{mnras}

\usepackage{newtxtext,newtxmath}

\usepackage[T1]{fontenc}

\DeclareRobustCommand{\VAN}[3]{#2}
\let\VANthebibliography\thebibliography
\def\thebibliography{\DeclareRobustCommand{\VAN}[3]{##3}\VANthebibliography}

\usepackage{graphicx}	
\usepackage{amsmath}	
\usepackage{multicol}  
\usepackage{pdflscape}	
\usepackage{hyperref}
\usepackage{siunitx}  
\newcommand{\GG}{\mbox{$G$}}
\newcommand{\GGp}{\mbox{$G^\prime$}}
\newcommand{\GGabs}{\mbox{$G_{\rm abs}$}}
\newcommand{\AGG}{\mbox{$A_G$}}
\newcommand{\GBP}{\mbox{$G_{\rm BP}$}}
\newcommand{\GRP}{\mbox{$G_{\rm RP}$}}
\newcommand{\dCC}{\mbox{$d_{\rm CC}$}}
\newcommand{\lili}{\mbox{LiLiMaRlin}}

\newcommand{\RV}{\mbox{$R_{5495}$}}

\newcommand{\Teff}{\mbox{$T_{\rm eff}$}}
\newcommand{\mci}[1]{\multicolumn{1}{c}{#1}}

\newcommand{\dOB}{\mbox{$d_{\rm OB}$}}
\newcommand{\dBJ}{\mbox{$d_{\rm BJ}$}}
\newcommand{\dSH}{\mbox{$d_{\rm SH}$}}

\title[A cross-match between the ALS catalog and Gaia DR2]{The Alma catalog of OB stars. II. A cross-match with Gaia DR2 and an updated map of the solar neighbourhood.}

\author[M. Pantaleoni Gonz\'alez et al.]{
M. Pantaleoni Gonz\'alez,$^{1, 2}$\thanks{E-mail: mpantaleoni@cab.inta-csic.es}
J. Ma{\'\i}z Apell\'aniz,$^{1}$
R. H. Barb\'a,$^{3}$
and B. Cameron Reed$^{4}$
\\
$^{1}$Centro de Astrobiolog{\'\i}a. CSIC-INTA. Campus ESAC. Camino bajo del castillo s/n. E-\num[detect-all]{28692} Villanueva de la Ca\~nada. Madrid. Spain.\\
$^{2}$Departamento de Astrof{\'\i}sica y F{\'\i}sica de la Atm\'osfera. Universidad Complutense de Madrid. E-\num[detect-all]{28040} Madrid. Spain.\\
$^{3}$Departamento de Astronom{\'\i}a. Universidad de La Serena. Av. Cisternas 1200 Norte. La Serena. Chile.\\
$^{4}$Department of Physics (Emeritus). Alma College. Alma. \num[detect-all]{48801} Michigan. United States of America.
}

\date{Accepted 2021 March 3. Received 2021 February 18; in original form 2020 November 30}

\pubyear{2020}

\begin{document}
\label{firstpage}
\pagerange{\pageref{firstpage}--\pageref{lastpage}}
\maketitle

\begin{abstract}
We cross-match the Alma catalog of OB stars with {\it Gaia}~DR2 astrometry and photometry as a first step towards producing a clean sample of massive stars in the solar neighbourhood with a high degree of completeness. We analyze the resulting colour-absolute magnitude diagram to divide our sample into categories and compare extinction estimates from two sources, finding problems with both of them. The distances obtained with three different priors are found to have few differences among them, indicating that {\it Gaia}~DR2 distances are robust. An analysis of the 3-D distribution of massive stars in the solar neighbourhood is presented. We show that a 
kinematically distinct
structure we dub the Cepheus spur extends from the Orion-Cygnus spiral arm towards the Perseus arm and is located above the Galactic mid-plane, likely being related to the recently discovered Radcliffe wave. We propose that this corrugation pattern in the Galactic disk may be responsible for the recent enhanced star formation at its crests and troughs. We also discuss our plans to extend this work in the immediate future.
\end{abstract}




\section{Introduction}

$\,\!$\indent The short lifespans of massive OB stars implies that they can be used as tracers of the large stellar formation regions and spiral arms of the Galaxy \citep{Morgetal52,Reed93b,BouyAlve15,Wardetal20}, with the exception of a small fraction of massive runaway stars \citep{Blaa93,Maizetal18b}. Going back to the studies of the 1950s, OB stars have been preferentially identified from spectroscopy. Photometric identifications can be problematic, especially for O stars, unless one has good-quality photometry to the left of the Balmer jump \citep{MaizSota08,Maizetal14a}, where the atmosphere complicates calibration for ground-based observations \citep{Maiz06a}. One solution to this problem is to leave the atmosphere behind and that is one of the reasons for the exquisite calibration of {\it Gaia} photometry \citep{Jordetal10}. However, the DR2 (2018) and EDR3 (2020) {\it Gaia} versions do not provide the full spectrophotometry that will become available in DR3 (expected for 2022) and, therefore, there is little information about the spectral energy distribution (SED) to the left of the Balmer jump in the current data (\citealt{MaizWeil18} and Appendix B here)

The ``Alma Luminous Star'' (ALS) catalog (\citealt{Reed03}, hereafter Paper I) is currently the largest published compilation of known Galactic luminous stars with available $UBV\beta$ photometric data. The ALS catalog was primarily built on the "Case-Hamburg" (C-S) survey for Galactic luminous stars, by joining data from \citet{StepSand71} in the south, with the northern component of the survey, published in six separate volumes \citep{Hardetal59,Stocetal60,NassStep63,Hardetal64,Hardetal65,Nassetal65}. Since its first appearance \citep{Reed93a}, the ALS catalog has grown to included more than 680 references. Since the publication of Paper I, over a decade and a half ago, a tremendous amount of information about OB stars has become available so this is an opportune moment to update the ALS catalog. In this second paper of the series we cross-match the sample from Paper I with the astrometric and photometric information from {\it Gaia}~DR2. In future papers of the series we will update the {\it Gaia} information with that of subsequent data releases and include data from other photometric and spectroscopic surveys such as spectra and uniform spectral classifications.

In the next section we describe our data and methods: how we have built the new version of the catalog and what its contents are. We then present our results about the colour-colour  and colour-absolute magnitude diagrams of OB stars, a comparison between different extinction and distance estimates, a study of the distribution of OB stars in the solar neighbourhood,
and an analysis of a new structure we dub the Cepheus spur. 
We end the paper with a description of our plans for future papers of the series.

\section{Data and methods}

\subsection{Building the new version of the catalog}

$\,\!$\indent Cross-matching a compilation of old sources with a modern and uniform database such as {\it Gaia}~DR2 is not straightforward, as there are different ways in which we can have false positives and false negatives: low-quality old coordinates, objects with large proper motions, duplicates, and erroneous identifications are some of them. To those, one has to add the cases where the star should not have been in the original catalogs in the first place because it did not really belong to its purported class of objects (which was unknown at the time). Therefore, there is a significant amount of detective work involved in the process if one desires a clean new version. A description of what we have done is provided in this subsection and details about the different types of problems are provided in Appendix~A.

We use here the traditional definition of an OB star: a massive star ($\gtrsim$~8~M$_\odot$) with an O or B spectral type. This corresponds to spectral types up to B2 for dwarfs, B5 for giants and B9 for supergiants. In principle, we exclude from the definition low-luminosity hot stars such as white dwarfs and subdwarfs, intermediate-type (AF) and late-type (GKM) supergiants, Wolf-Rayet stars, and others. Also, as we intend the catalog to be a Galactic one we plan to separate extragalactic objects from the main sample. Of course, excluding from the definition is not the same as excluding from the sample, the latter being more difficult due to the diverse quality of the data. For that reason, we will proceed with caution on this series as to how we determine what each object is, placing them in this first paper in temporary categories based on {\it Gaia}~DR2 photometric and astrometric information alone and leaving for subsequent papers a permanent assignment based on further spectroscopic and photometric data. Also, objects determined to be e.g. A supergiants will not be completely excluded from the catalog in future papers but instead placed on appropriate supplements. Our goal in this first paper then is to produce a clean version of the ALS catalog with good-quality photometric and astrometric information and where we can use it to produce a preliminary classification of the objects into likely massive stars and other types. 

We start the cleaning process with the current version of the ALS catalog (updated in 2005), which contains \num{18693} entries (Table~\ref{stats}). Most of the objects correspond to OB stars but $\sim$10\% of the entries correspond to other types of objects, which paper~I indicates they are mostly A-G supergiants, white dwarfs (WDs), planetary nebula nuclei (PNNi), and Wolf-Rayet stars (WRs). \num{393} of these are duplicates: 298 flagged as so in the ALS catalog itself, 72 matched by the CDS on Simbad's set of identifiers, and 23 additional duplicates found by us in this work. Not all duplicates are pairs of identifiers, the term "duplicate" here includes 6 triplets in the original catalog. We remove the duplicates from the new version of the catalog and flag them with a D, leaving the number of unique objects at \num{18300}. These 393 duplicates do not always reflect identical instances of the same object, as they are drawn from different parts of the literature and can also appear in situations where modern references account for multiple objects (or a particular component of a multiple system) while older references could not resolve the parts of the system, in which case we consider the identifier linked to the old reference as a duplicate of the brightest component in the set of resolved sources. Multiplicity, spectroscopic or visual, is an ubiquitous characteristic of massive stars \citep{Masoetal98,Sotaetal14,Maizetal19b} and this leads to the question of how to name multiple components in the catalog (with independent entries or with A,B,C\ldots extensions). We will not address it in this paper but will do so in future articles in this series.

\begin{table}
\caption{Number of objects by category in this paper. The first block details the cleaning process until arriving to the final sample of \num{15662} stars and the second block the breakdown of the final sample by categories.}
\label{stats}
\begin{tabular}{clr}
Cat. & Description & \mci{\#} \\
\hline
  & Objects in the 2005 version of the ALS catalog        & \num{18693} \\
  & After eliminating 393 duplicates (D)                  & \num{18300} \\
  & After eliminating 211 unmatched objects (U)           & \num{18089} \\
 & After eliminating 2336 stars with bad astrometry (A)   & \num{15753} \\
 & After eliminating 91 objects with bad colours (C)      & \num{15662} \\
\hline
M & Likely massive stars                                  & \num{13762} \\
I & High/intermediate-mass stars                          & \num{1506}  \\
L & Intermediate/low-mass stars                           & \num{260}   \\
H & High-gravity stars                                    & \num{127}   \\
E & Extragalactic stars                                   & \num{7}     \\
\hline
\end{tabular}
\end{table}

After eliminating duplicates, we proceed to cross-match the ALS catalog with {\it Gaia}~DR2 sources. Currently, all ALS identifiers can be queried in Simbad with the exception of ALS~1823, ALS~\num{12636}, ALS~\num{15196}, ALS~\num{15858} and ALS~\num{19764}, which are recognized in Simbad as CD~$-$58~3529, BD~$+$61~2352, Trumpler~14~9, CPD~$-$58~2655, and LS~4723, respectively. Including those five objects by hand, \num{16912} of the \num{18300} unique ALS entries are currently cross-matched in {\it Gaia}~DR2 by Simbad. For the remaining 1388 entries we search for {\it Gaia} DR1, Hipparcos, Tycho, 2MASS, and WISE identifiers in Simbad and then we use the crossmatch products of the {\it Gaia} Archive to retrieve a {\it Gaia} DR2 source identifier. This can be successfully done for an additional 684 ALS identifiers. Finally, we address the remaining 704 unmatched cases manually by examining the available references attached to the ALS and by searching for astrophotometrically similar sources within a few arcseconds and small magnitude differences (the margins of these criteria are a function of  crowding). To accomplish those matches we used Aladin and for high proper motion objects the back-propagation of the trajectories was executed before the separation threshold was applied. In this way, 493 additional ALS sources were successfully matched with {\it Gaia}~DR2. This leaves us with only 211 unique unmatched sources and a total of \num{18089} matched ones ($98.8\%$ of the unique sources). 

Why are there 211 unmatched ALS sources in {\it Gaia}~DR2? There are different reasons. Some are too bright in \GG\ to be included in {\it Gaia}~DR2 (e.g. ALS~\num{14793}~=~$\zeta$~Ori~A) and some are too faint (e.g. ALS~\num{19589}). Others are in crowded regions like NGC~3603, where in some cases it is unclear which source the original paper refers to. We list those objects in a supplemental table but leave them out of our final sample for this paper. It is our plan to incorporate at least some of them into the catalog in future papers.

The next step is the astrometric quality control. There are 2343 remaining objects with either (a) no {\it Gaia}~DR2 parallaxes ($\varpi$), (b) large uncertainties in the parallax ($\varpi/\sigma_\varpi < 3$), or (c) bad astrometric solutions (RUWE~$>$~1.4). We eliminate all of those from the sample except for seven which are located in the Magellanic Clouds (see below)  to arrive at a number of \num{15753}~objects.

\begin{figure*}
\centerline{
\includegraphics[width=0.49\linewidth, trim=0.5cm 0.5cm 0.0cm 0.5cm]{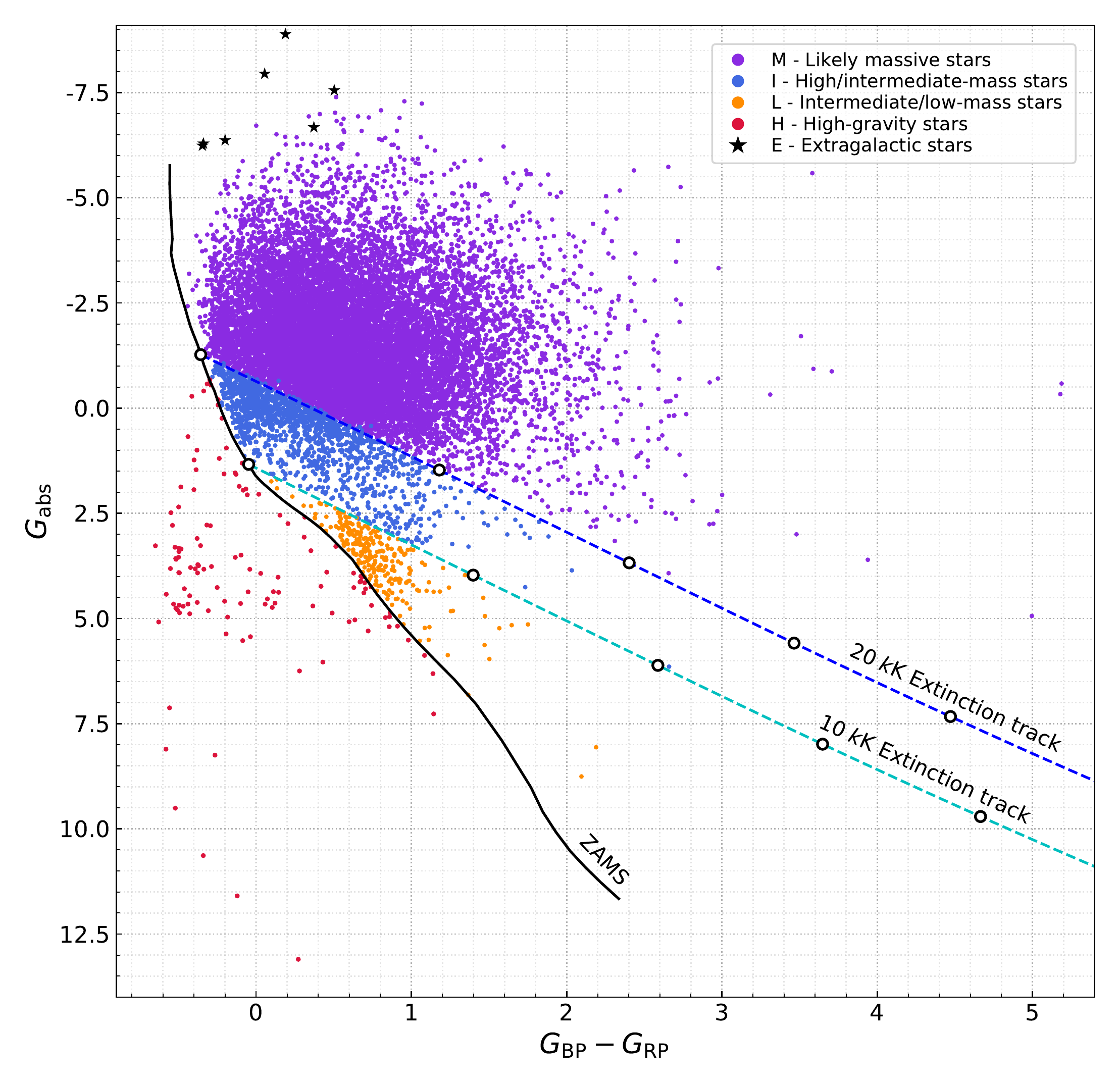} \ 
\includegraphics[width=0.49\linewidth, trim=0.5cm 0.5cm 0.0cm 0.5cm]{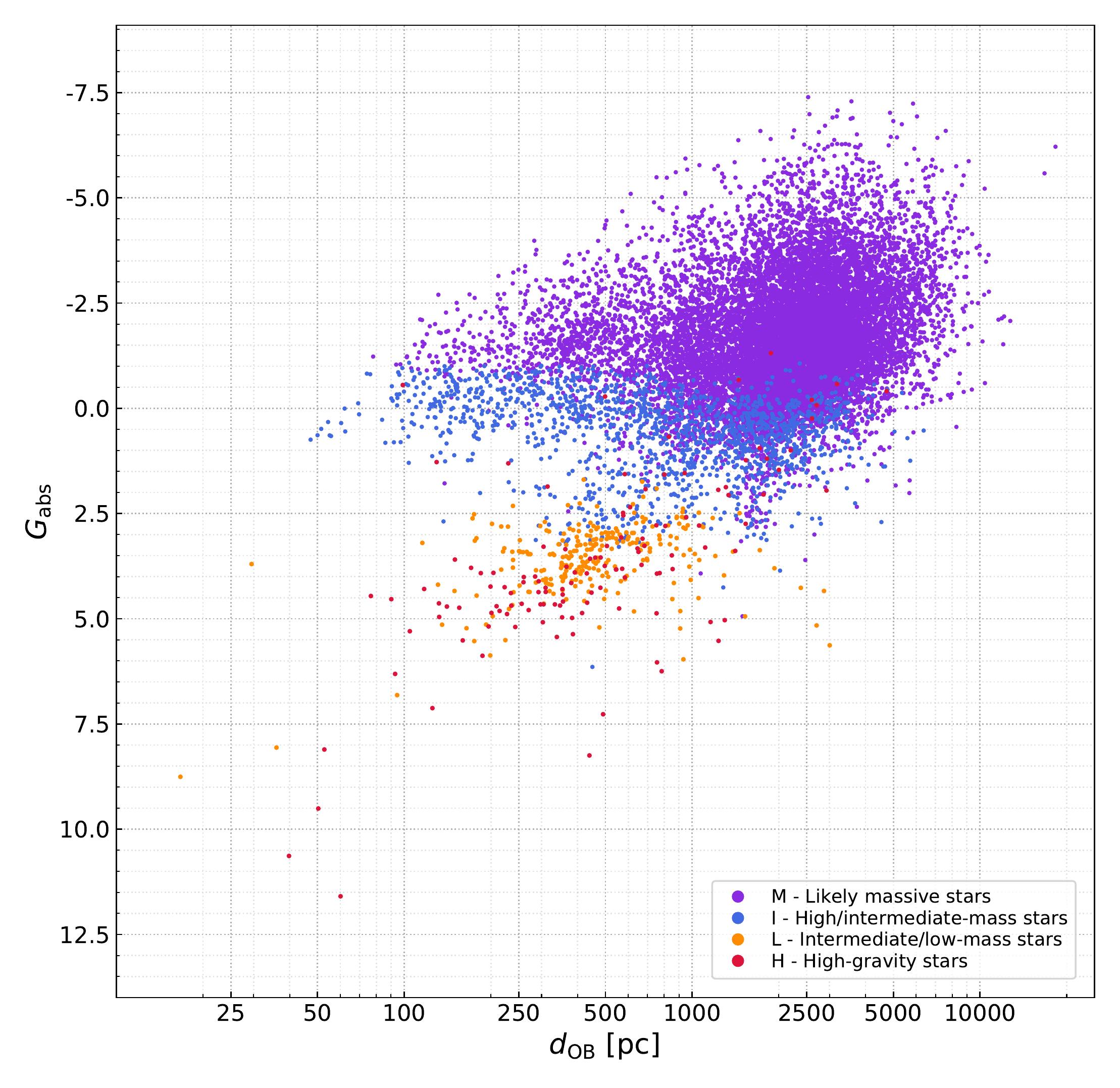}
}
\caption{(left) {\it Gaia}~DR2 \GBP$-$\GGp~+~\GGabs\ colour-absolute magnitude diagram for the final sample in this paper. The sample is colour-coded by category. The three boundaries used to divide the sample in the four Galactic categories are shown. (right) Distance-absolute magnitude diagram for the same sample without the seven extragalactic stars and ALS~\num{2481} (=~Gliese~440, a white dwarf at a distance of 4.6~pc). Markings on the extinction curves correspond to $E(4405-5495) = 0$, $1$, $2$, $3$ and $4$ magnitudes.}
\label{colmabs}
\end{figure*}

The last step to arrive at the final sample is the photometric quality control, that is, the elimination of targets with incomplete (one or more missing)
or bad-quality {\it Gaia}~DR2 \GBP+\GG+\GRP\ magnitudes. With respect to what we consider as bad-quality photometry, \GBP\ and \GRP\ magnitudes are expected to be contaminated in crowded fields and bright sources may be saturated in one or more of the three bands. Those two are the reasons we have to apply this step, for which we selected the criterion $|\dCC| > 0.15$ and eliminated a total of 91 objects with bad colours. \dCC\ is the distance from the stellar locus in the \GBP$-$\GGp~+~\GGp$-$\GRP\ plane and \GGp\ is the {\it Gaia}~DR2 corrected \GG\ magnitude \citep{MaizWeil18,Maiz19}. In future papers, with information from other sources and new {\it Gaia} data releases, most of these objects should be reintegrated into the main part of the catalog.  

Having arrived at our final sample, we calculate \GGabs\ using the distance \dOB\ described below in subsection~\ref{distanceestimates}. \GGabs\ is computed from \GGp\ but is not corrected from extinction. Exceptions for the distance are made for seven extragalactic objects in the Magellanic Clouds that were inadvertently included in the original catalog (ALS~\num{15895}, ALS~\num{15896}, ALS~\num{18185}, ALS~\num{18840}, ALS~\num{18845}, ALS~\num{19597}, and ALS~\num{19598}), for which we use distances of 62~kpc (SMC, ALS~\num{19597}) and 50~kpc (LMC, rest of the sample). We use the $\GBP-\GRP$ vs. $G_{\rm abs}$ colour-absolute magnitude diagram in Fig.~\ref{colmabs} to classify the final sample into five categories: likely massive stars (M), objects in the high/intermediate mass regime (I), objects that are likely to be intermediate- or low-mass stars (L), high-gravity stars (H), and extragalactic objects (E). The four Galactic categories are defined using as boundaries the ZAMS 
from the solar metallicity grid of \citet{Maiz13a}
and the extinction tracks for ZAMS stars with 10~kK and 20~kK with the $\RV=3.0$ extinction law of \citet{Maizetal14a}. The $\RV=3.0$ value is chosen from \citet{MaizBarb18} as representative of the typical intermediate-to-high extinction for OB stars (note that in the low-extinction case higher values are more common). The ZAMS separates high-gravity stars (white dwarfs and subdwarfs) from the rest of the sample. The values for the \Teff\ of the extinction tracks are selected to represent the approximate value for the A0~V and B2.5~V spectral subtypes in the MS, respectively. Note that evolved OB stars such as intermediate-B giants and later-type supergiants are also located above the 20~kK track independent of their extinction due to their intrinsic \GGabs\ values.  
We repeat here that these five are temporary categories and that the membership to each one of them will be reanalyzed in subsequent papers. For example, some type L stars may be extinguished H-type objects. For the purposes of this paper we will use the type M stars as our ``OB massive star sample'' but we remind the reader that it has a few contaminants in the form of later-type supergiants. 

\subsection{Catalog description}

\begin{table*}
\caption{Description of the contents in the table sent to the CDS with the final sample.}
\label{cds}
\centerline{
\begin{tabular}{lcl}
Column name       & Units    & Description                                                          \\
\hline
\multicolumn{3}{c}{Category and identifiers}                                                        \\
\hline
Cat               & ---      & Category (M/I/L/H/E)                                                 \\
ID\_ALS           & ---      & Alma Luminous Star number                                            \\
ID\_DR2           & ---      & {\it Gaia} DR2 source identifier                                     \\
ID\_GOSC          & ---      & Galactic O-Star Catalog identifier                                   \\
\hline
\multicolumn{3}{c}{Coordinates}                                                                     \\
\hline
RA\_ALS           & hh:mm:ss & Right ascension (J2000.0) from original ALS catalog                  \\
DEC\_ALS          & dd:mm:ss & Declination (J2000.0) from original ALS catalog                      \\
RA\_DR2           & hh:mm:ss & Right ascension (Epoch 2015.5) from {\it Gaia} DR2                   \\
DEC\_DR2          & dd:mm:ss & Declination (Epoch 2015.5) from {\it Gaia} DR2                       \\
GLON              & deg      & Galactic longitude from {\it Gaia} DR2                               \\
GLAT              & deg      & Galactic latitude from {\it Gaia} DR2                                \\
\hline
\multicolumn{3}{c}{Photometry}                                                                      \\
\hline
Vmag              & mag      & $V$ magnitude from original ALS catalog                              \\
Gmag              & mag      & \GG\ magnitude in {\it Gaia} DR2                                     \\
Gmag\_cor         & mag      & \GGp, corrected \GG\ magnitude                                       \\
BPmag             & mag      & \GBP\ magnitude in {\it Gaia} DR2                                    \\
RPmag             & mag      & \GRP\ magnitude in {\it Gaia} DR2                                    \\
BPmag-Gmag\_cor   & mag      & \GBP$-$\GGp\ colour                                                  \\
Gmag\_cor-RPmag   & mag      & \GGp$-$\GRP\ colour                                                  \\
\hline
\multicolumn{3}{c}{Quality indicators}                                                              \\
\hline
CCDIST            & mag      & \dCC, distance to the {\it Gaia} DR2 colour-colour main locus        \\
Sep\_astr         & arcsec   & Separation between ALS and {\it Gaia} DR2 positions                  \\
RUWE              & ---      & Renormalised Unit Weight Error in Gaia DR2                           \\
\hline
\multicolumn{3}{c}{Proper motions}                                                                  \\
\hline
PM\_RA            & mas/a    & Proper motion in right ascension in {\it Gaia} DR2                   \\
PM\_RA\_err       & mas/a    & Proper motion uncertainty in right ascension in {\it Gaia} DR2       \\
PM\_DEC           & mas/a    & Proper motion in declination in {\it Gaia} DR2                       \\
PM\_DEC\_err      & mas/a    & Proper motion uncertainty in declination in {\it Gaia} DR2           \\
\hline
\multicolumn{3}{c}{Parallax and distance}                                                           \\
\hline
Plx               & mas      & $\varpi$, parallax in {\it Gaia} DR2                                 \\
Plx\_err          & mas      & $\sigma_\varpi$, parallax uncertainty in {\it Gaia} DR2              \\
ALS2\_dist\_mode  & pc       & Mode of the posterior distribution of distances                      \\
ALS2\_dist\_P50   & pc       & \dOB, median of the posterior distribution of distances              \\
ALS2\_dist\_mean  & pc       & Mean of the posterior distribution of distances                      \\
ALS2\_dist\_HDIl  & pc       & Lower limit of the highest density interval for $p$ = 68\%           \\
ALS2\_dist\_HDIh  & pc       & Upper limit of the highest density interval for $p$ = 68\%           \\
ALS2\_dist\_P16   & pc       & 16th percentile of the posterior distribution of distances           \\
ALS2\_dist\_P84   & pc       & 84th percentile of the posterior distribution of distances           \\
BJ\_dist          & pc       & \dBJ, Bailer-Jones distance estimate                                 \\
BJ\_dist\_l       & pc       & Bailer-Jones lower distance estimate                                 \\
BJ\_dist\_h       & pc       & Bailer-Jones upper distance estimate                                 \\
SH\_dist          & pc       & \dSH, StarHorse distance estimate                                    \\
SH\_dist\_l       & pc       & StarHorse lower distance estimate                                    \\
SH\_dist\_h       & pc       & StarHorse upper distance estimate                                    \\
\hline
\multicolumn{3}{c}{Absolute magnitude}                                                              \\
\hline
Gabs\_mode        & mag      & \GGabs\ from the mode of the posterior as a distance estimate        \\
Gabs\_mode\_err   & mag      & \GGabs\ uncertainty from the highest density interval for $p$ = 68\% \\
Gabs\_P50         & mag      & \GGabs\ from the median of the posterior as a distance estimate      \\
Gabs\_P1684       & mag      & \GGabs\ uncertainty from the 16th and 84th percentiles               \\
\hline
\multicolumn{3}{c}{Other information}                                                               \\
\hline
SpT\_ALS          & ---      & Spectral classifications from original ALS catalog                   \\
Sflag             & ---      & Simbad cross-match flag (N/S/W)                                      \\
Pflag             & ---      & $V$ magnitude type from original ALS catalog (N/P/V)                 \\
Other\_crossmatch & ---      & Other {\it Gaia} DR2 cross-match candidates                          \\
Comments          & ---      & Notes                                                                \\
\hline
\end{tabular}
}
\end{table*}

\begin{figure*}
\centerline{
\includegraphics[width=0.49\linewidth, trim=0.5cm 0.5cm 0.0cm 0.5cm]{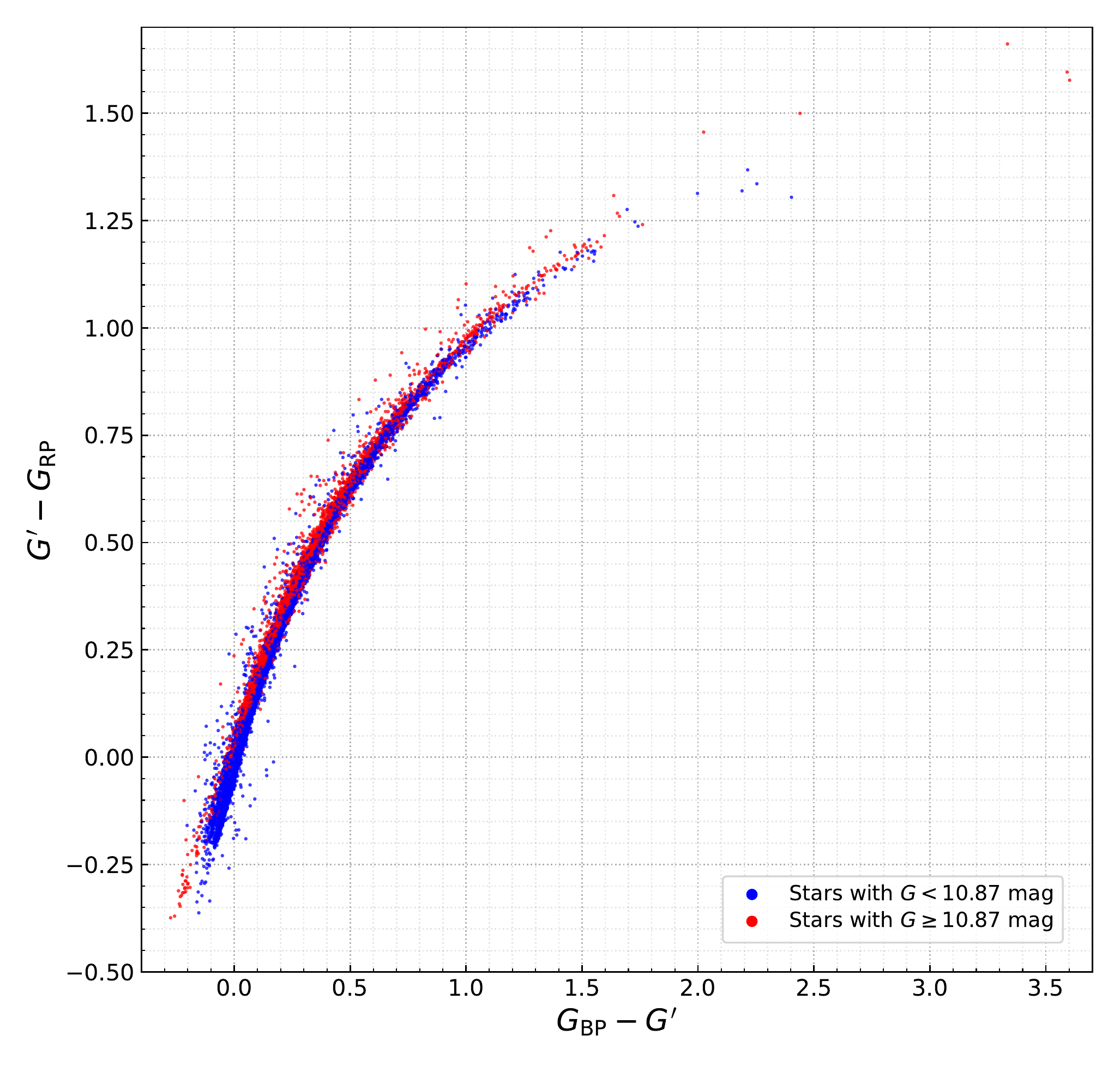} \ \includegraphics[width=0.49\linewidth, trim=0.5cm 0.5cm 0.0cm 0.5cm]{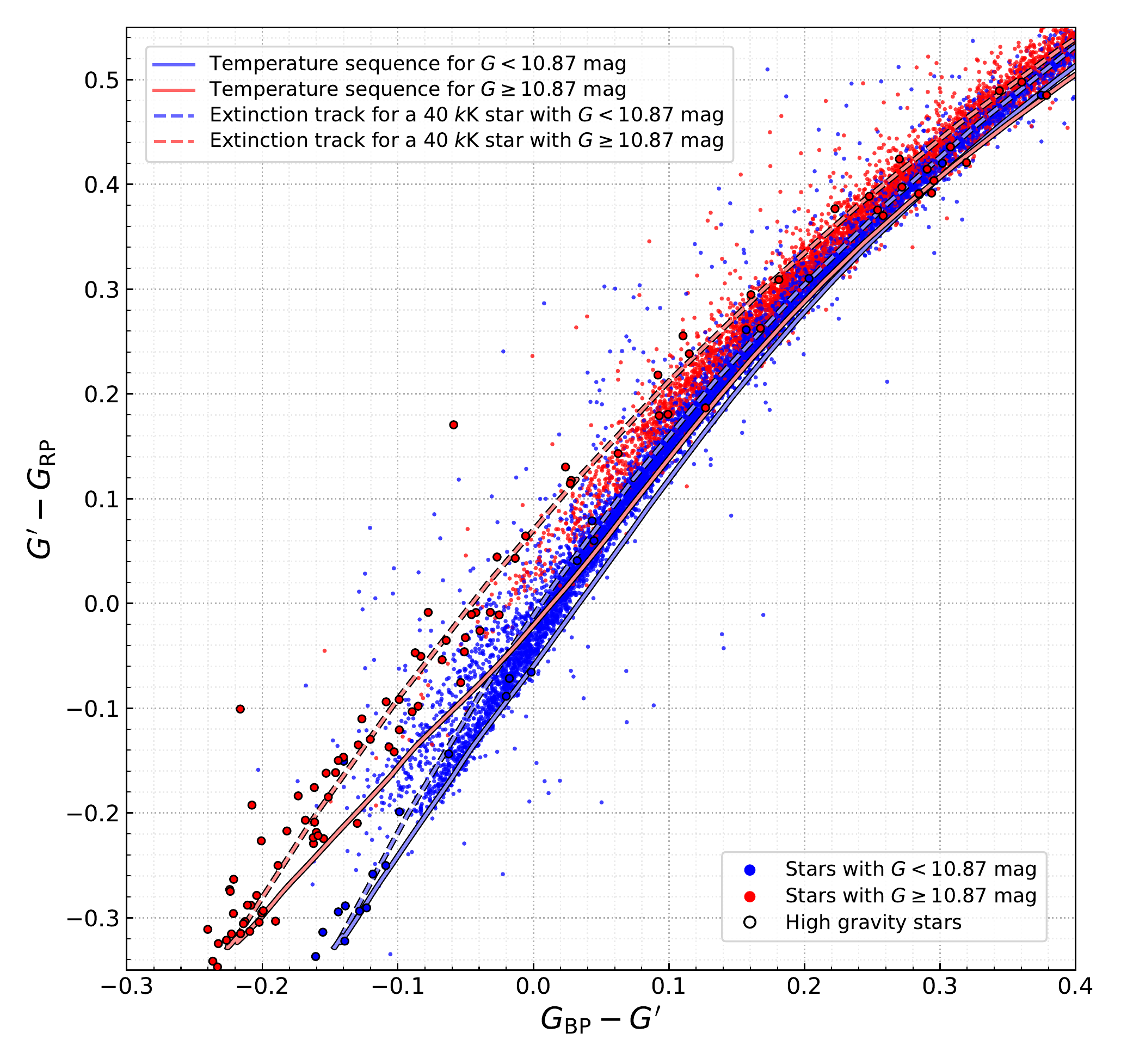}
}
\caption{{\it Gaia}~DR2 \GBP$-$\GGp~+~\GGp$-$\GRP\ colour-colour diagram for the final sample in this paper. The points are colour-coded depending on their value of \GG\ (bright or faint stars). The left panel shows the full range spanned by the sample while the right panel is a zoom into the low-extinction region. In the right panel the high-gravity subsample is shown with additional black circumferences and the lines show the (a) \Teff\ MS zero-extinction sequences and (b) the \RV~=~3.0 extinction tracks for a MS star with \Teff~=~40~kK for both the bright and faint magnitude ranges using the calibration of \citet{MaizWeil18}.}
\label{colcol}
\end{figure*}

$\,\!$\indent The result of the cleaning and classification process are presented in the form of a main catalog and three supplements. The main catalog lists our final sample of \num{15662} stars and the three supplements lists the \num{2427} stars excluded due to bad astrometry or colours, the \num{211} unmatched objects, and the \num{393} duplicates. The information in the main catalog is described in Table~\ref{cds}. The supplements give the same information as the catalog whenever applicable. The catalog and supplements are available in electronic format only due to the large number of columns.

The codes used for the category in Table~\ref{cds} are the same as those in Table~\ref{stats}. The codes for the Sflag indicate if Simbad successfully matches the ALS identifier with the correct {\it Gaia}~DR2 identifier (S), if the match is different from ours (W), or if it did not perform a match at all (N). The codes for the Pflag indicate the photometric type in the ALS: V when the magnitude was retrieved from photoelectric measurements, P when it was obtained from photographic plates, and N when the original ALS lacked photometry.

\section{Results}

\subsection{The colour-colour and colour-absolute magnitude diagrams}

$\,\!$\indent We plot in the left panel of Fig.~\ref{colcol} the \GBP$-$\GGp~+~\GGp$-$\GRP\ colour-colour diagram for the full clean sample of \num{15662} stars in the paper, colour-coded by their membership to the bright or faint magnitude ranges. The intrinsic colours of the vast majority of the sample are negative (lower-left part of the plot) and the curved sequence we see in the plot is for the most part an extinction sequence. The curvature is caused by the need to integrate over very broad band filters to accurately calculate extinction. If non-linear extinction effects are ignored (e.g. by assuming that $A_G$ is proportional to $E(\GBP-\GRP)$), significant biases can be introduced in the result \citep{Maiz13b,MaizBarb18,Maizetal20a}. Furthermore, extinction not only reddens colours but also makes stars fainter overall (again, non-linearly, the extinction tracks in Fig~\ref{colmabs} are slightly curved). For that reason, most stars in the lower left part of the plot are bright while faint stars concentrate in the central and upper right regions (of course, there are also selection effects involved). The right panel of Fig.~\ref{colcol} is a zoom into the bottom left region of the left panel, where all OB stars should be if it were not for extinction, and the differences seen there as a function of magnitude range are a {\it Gaia}~DR2 calibration issue, as described in Appendix~B.

As already mentioned, the colour-absolute magnitude diagram in Fig.~\ref{colmabs} is used to classify the Galactic part of the sample into four categories, with seven additional objects being located in the Magellanic Clouds. The numbers in each category are given in Table~\ref{stats}. As expected, the vast majority (\num{13762}/\num{15662} or 87.9\%) are classified as type M and most of the rest (9.6\%) are of type I, that is, objects near the real boundary between the two categories. Only 2.5\% of the sample is clearly excluded, indicating that the original ALS catalog was a relatively clean sample but not a perfect one, especially considering that our procedure does not allow us to discriminate between OB stars and later-type supergiants, which are known to be contaminants.

The spread on the M- (and I-) type objects in the horizontal direction in Fig.~\ref{colmabs} is due to a combination of two effects. The first one is the spread in intrinsic colours caused by the different values of \Teff\ for OB stars (plus the later-type-supergiant contaminants). However, A0 stars have $\GBP-\GRP \sim 0.0$ so this effect is a minor one. The second effect is the largest one and is extinction, which is analyzed in more detail in the next subsection. 

Another feature seen in the colour-absolute magnitude diagram is the different behaviour of the I- and M-type objects near the main sequence. Intermediate-mass stars are seen relatively close to the ZAMS with a small gap caused by the stars experiencing small amounts of extinction or evolution from their zero age. As we move to M-type objects the gap widens significantly as a combination of four effects: (a) As seen in the right panel of Fig~\ref{colmabs}, close to the Sun I-type objects are more abundant than M-type ones, thus increasing the chances of finding more low-extinction I-types than M-types. (b) Stars near the top of the diagram are more likely to be B supergiants than extinguished O stars because the bolometric correction makes them intrinsically brighter in \GG\ at constant luminosity. (c) The IMF and the MS lifetimes decrease as we move upwards in the diagram. (d) Finally, O-type stars near the ZAMS are hard to find anywhere in the Galaxy, especially the earlier subtypes \citep{Holgetal20}. These four effects will be discussed in future installments of this series.

\begin{figure*}
\centerline{
\includegraphics[width=0.49\linewidth, trim=0.5cm 0.5cm 0.0cm 0.5cm]{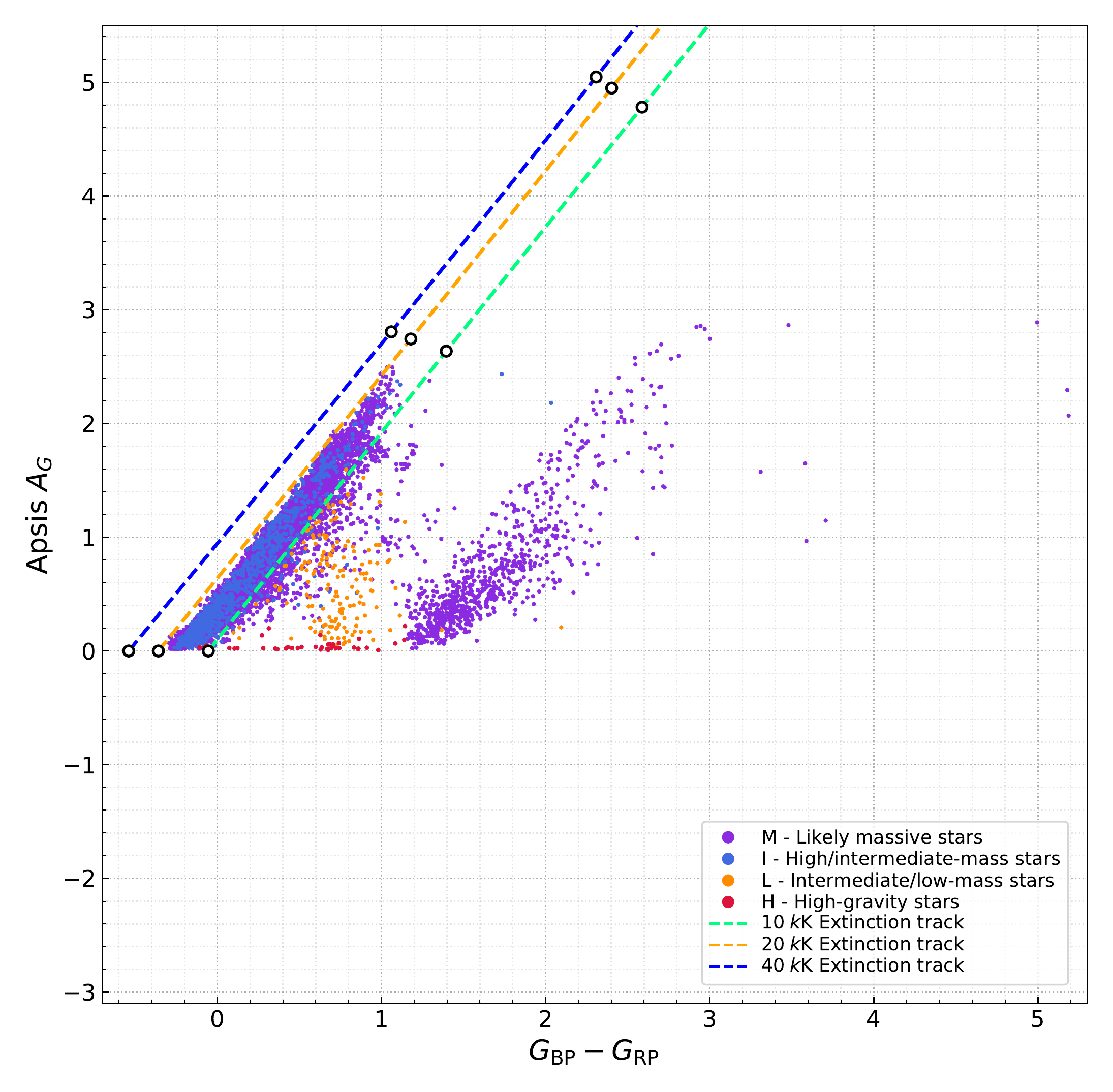} \ 
\includegraphics[width=0.49\linewidth, trim=0.5cm 0.5cm 0.0cm 0.5cm]{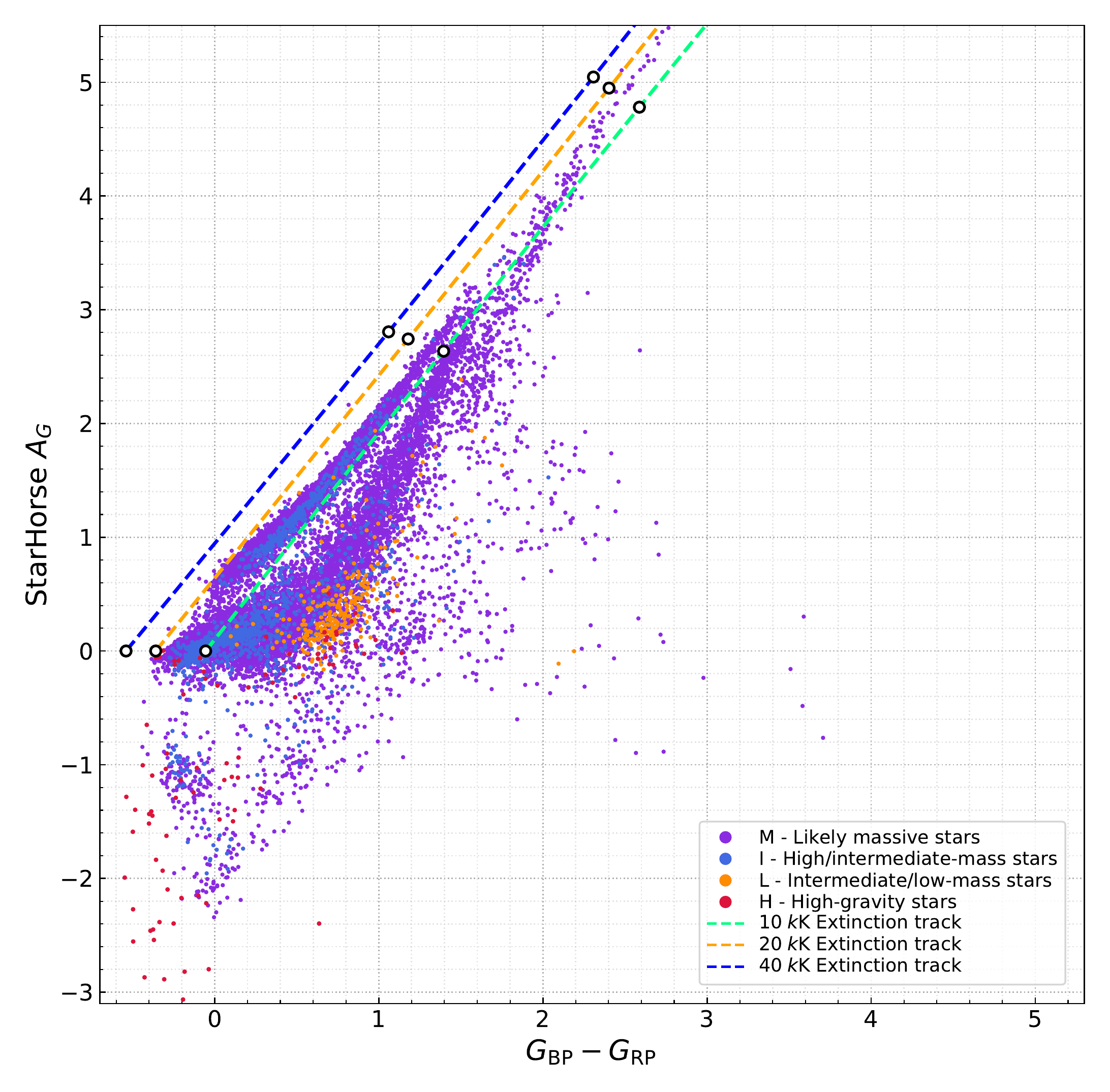}
}
\caption{Comparison between the values for the extinction \AGG\ as a function of \GBP$-$\GRP\ obtained using Apsis (left) and StarHorse (right) for the final sample. In both cases we also plot the expected relationship between the two quantities for MS stars of 10~kK, 20-kK, and 40~kK and a extinction law from \citet{Maizetal14a} with \RV~=~3.0. Markings on the extinction curves correspond to $E(4405-5495) = 0$, $1$ and $2$ magnitudes.}
\label{extcomp}
\end{figure*}

The position of the seven MC objects in the colour-absolute magnitude diagram also deserves comment. Four of them are among the brightest objects in \GGabs, which we remind the reader is not extinction-corrected, and the other three are in the gap close to the ZAMS described in the previous paragraph. In both cases the low extinction of most stars in the Magellanic Clouds compared to the OB Galactic sample plays a role in placing those stars towards the upper left corner of the diagram. In the case of the first four objects a second explanation is that they are of spectral type BIa, i.e. very luminous supergiants with small bolometric corrections in the \GG\ band. In the case of the last three objects, they are two early O-stars and an O+O binary, placing them close to the extinction-free leftmost possible location in the colour-absolute magnitude diagram.

Finally, it is instructive to compare our colour-absolute magnitude diagram with Fig.~1 of \citet{Babuetal18}, which is the equivalent diagram using the \num{65921112} stars with good-quality {\it Gaia}~DR2 data independently of their spectral types. As a first-order approximation, both diagrams are complementary: the region with a high-density of objects in Fig.~\ref{colmabs} here appears as a low-density region in the diagram of the general {\it Gaia}~DR2 population, with most stars there located towards the right and bottom. This is a manifestation of what dominates the {\it Gaia}~DR2 population: low-mass stars near the main sequence and red giants (most noticeably in the diagram, red-clump stars, clearly seen as an extinction sequence). This in turn is a consequence of the very large numbers of the first type and of the right combination of high (but not as large as the first) numbers and high luminosities for the second type. However, a second, more subtle effect is present in the comparison between the two diagrams. For luminous stars the transition from low-density to high-density is rather abrupt in the \citet{Babuetal18} around $\GBP-\GRP = 1.0-1.2$ due to the appearance of the first low-extinction red-clump stars but why should that be accompanied by a complementary transition among the M+I type stars in our diagram? The reason is that OB stars are relatively easy to identify by their colours as long as they are not too extinguished. When they are, they of course become fainter but being intrinsically luminous objects they should still be detected. The real problem is that around $\GBP-\GRP = 1.0-1.2$ (or their equivalent in other photometric systems) the much more numerous population of Galactic red giants makes their appearance and severely hampers the identification of OB stars. That is why compilations of relatively old catalogs like the previous version of the ALS have only a few high-extinction objects. Detecting OB stars in the Galactic plane is like finding a needle in a haystack.

\subsection{Comparing extinction estimates}

$\,\!$\indent As discussed in the previous subsection, most of the spread in the horizontal direction in Fig.~\ref{colmabs} is caused by extinction. We have previously measured the extinction properties of part of the ALS sample \citep{MaizBarb18,Maizetal21a} but only for a small fraction of it, so we will not consider those results here and we will leave our extinction analysis for a future installment of this series. Instead, here we analyze the extinction estimates in the \GG\ band (\AGG) by \citet{Andretal18} using Apsis \citep{Bailetal13} from {\it Gaia}~DR2 alone and by \citet{Andeetal19} using StarHorse \citep{Queietal18} combining {\it Gaia}~DR2 information with Pan-STARRS1, 2MASS, and WISE. We have cross-matched our sample with both of those results and found numbers of \num{8701} for Apsis and \num{13844} for StarHorse, with \num{7799} objects in common between the three papers. We plot in Fig.~\ref{extcomp} \AGG\ as a function of \GBP$-$\GRP\ for both techniques.

\begin{figure*}
\centerline{
\includegraphics[width=0.49\linewidth, trim=0.5cm 0.5cm 0.0cm 0.5cm]{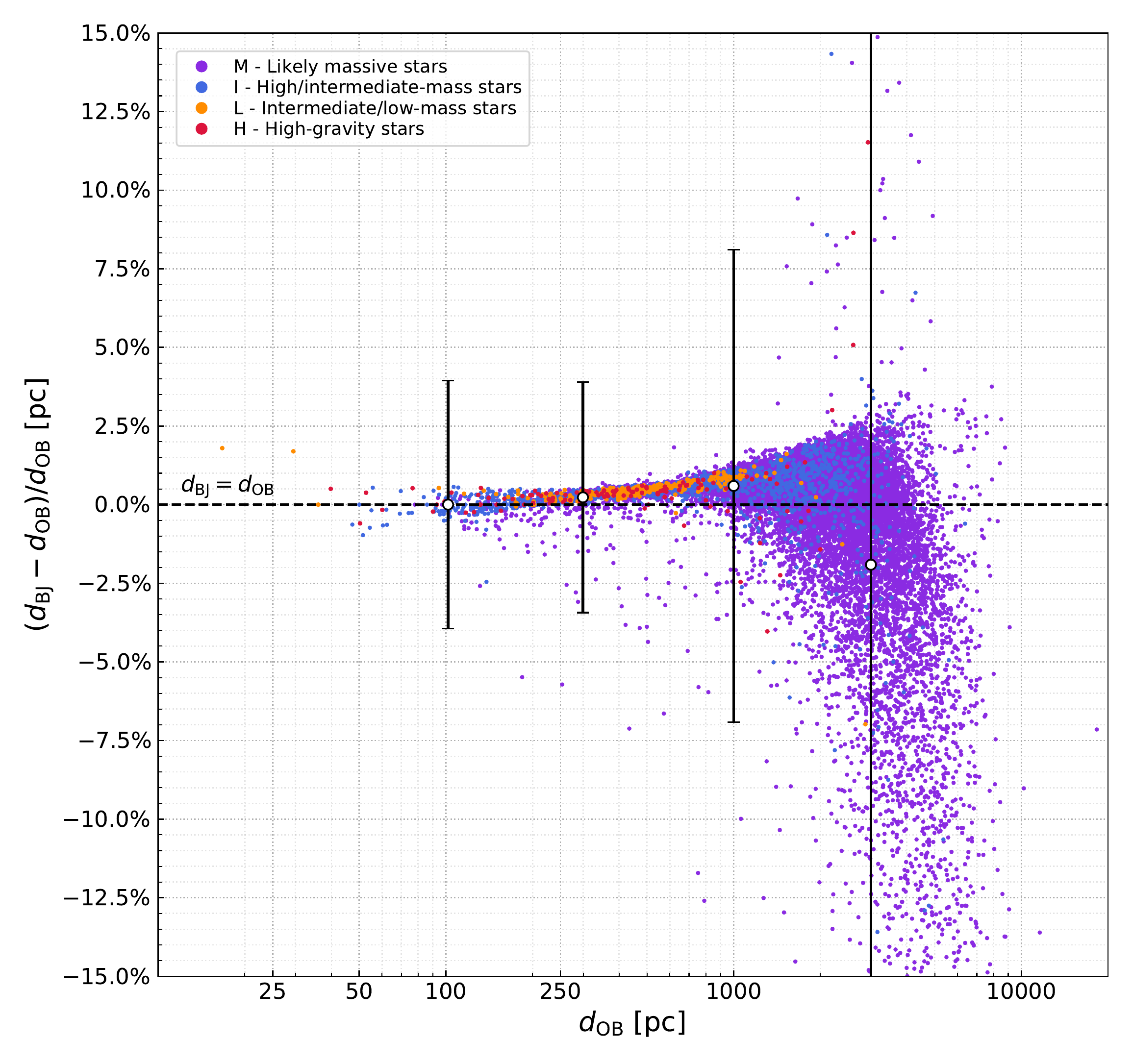} \ 
\includegraphics[width=0.49\linewidth, trim=0.5cm 0.5cm 0.0cm 0.5cm]{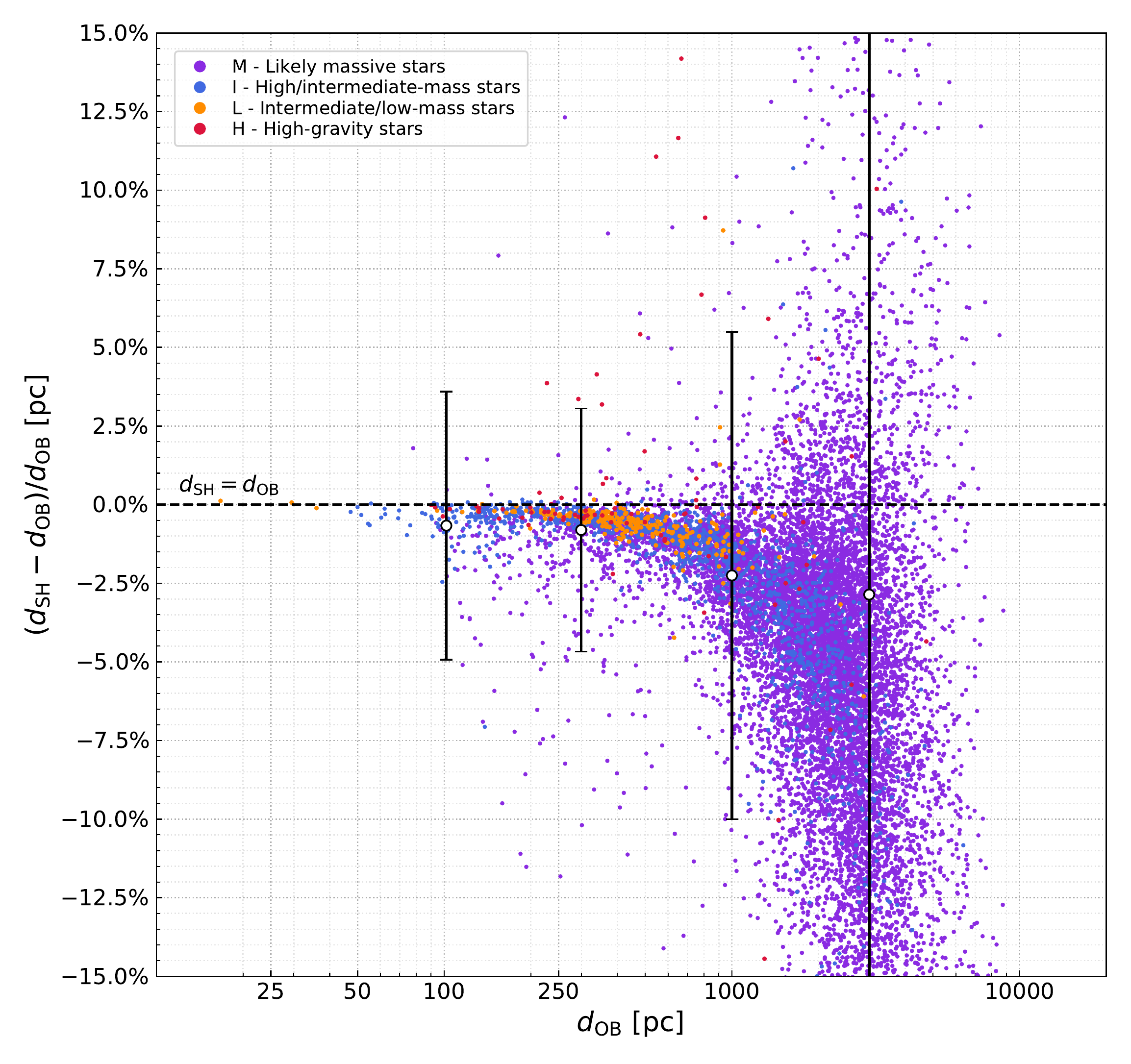}
}
\caption{Comparison between distances obtained using the OB and the Bailer-Jones priors (left) and between the OB and the StarHorse priors (right) for the final ALS sample colour-coded as in Fig.~\ref{colmabs}. The comparison is made between the medians of the posterior distribution. The error bars, on the other hand, correspond to one standard deviation of the posterior distribution for typical stars at 100, 300, 1000, and 3000~pc, respectively. The error bars for 3000 kpc are larger than the plotted range and amount to 18\% (left panel) and 21\% (right panel).}
\label{distcomp}
\end{figure*}

We first analyze the Apsis results in Fig.~\ref{extcomp}. On the positive side, there is a linear trend between \GBP$-$\GRP\ and \AGG\ for the bulk of the stars, indicating that the non-linearity effect of extinction is correctly accounted for. Also, high-gravity stars have low extinctions, stars of type L are in the region of the plot where they are supposed to be, and there are no objects with negative extinctions (a condition expressly imposed by \citealt{Andretal18}). On the negative side, \citealt{Andretal18} trained their algorithm using models up to 20~kK for calculating extinction (and up to 10~kK for their calculation of \Teff) and, as a consequence, the bulk of their stars is contained within the relationships for 10~kK and 20~kK, while most of the stars there actually have values of \Teff\ between 20~kK and 40~kK. Therefore, the \citealt{Andretal18} \AGG\ values are consistently underestimated for most of the M-type stars (and also possibly for I-type stars) with \GBP$-$\GRP$\lesssim$1.1 in our sample by several tenths of a magnitude. A corollary of this is that the bluemost objects in the sample have no Apsis extinctions, as they would require negative values of \AGG\ that the algorithm does not allow. Another problematic aspect is that the main trend (located in the region expected for stars with \Teff\ in the range 10-20~kK) stops around \GBP$-$\GRP$\sim$1.1 and continues in a parallel track that goes back to zero extinction starting at that same colour and containing about 800~stars. We have verified that a few of those 800~stars are some of the later-type supergiants still present among the M-type stars but the majority of them are bona fide OB stars, including a significant number of O~stars with accurate GOSSS \citep{Maizetal11} spectral types. The likely explanation for this effect is that the Apsis algorithm considers that the vast majority of objects with \GBP$-$\GRP$\gtrsim$1.1 cannot be OB stars and are instead assigned a low value of \Teff\ and, hence, a lower value of \AGG.

The structures seen in the StarHorse plot of Fig~\ref{extcomp} are very different from those seen in the Apsis plot. In the first place, there are 1502 stars (10.8\% of the total) with negative extinction values (non-negativity was not imposed as a condition) extending to values of $\AGG\sim-3$~mag, indicating that at least some OB stars were erroneously identified by the algorithm as being of later spectral type. On the other extreme, we find \AGG\ values of more than $\sim$5~mag for stars with estimated \Teff\ above 10~kK. This is as expected but we note such extinction values are missing on the Apsis results (see above). In between the two extinction extremes, two different trends are seen in the \AGG~=~0-3~mag, one above the other and both with a curvature. The origin of the curvature is likely to be an inaccurate treatment of non-linear extinction effects. As it happened with the Apsis results, there are no stars in the region of the diagram expected for 20-40~kK stars, leading us to suspect that \AGG\ is underestimated for most of the stars.

In summary, both \citet{Andretal18} and \citet{Andeetal19} \AGG\ estimates are not optimized for OB stars and for the ALS sample they are in general underestimates and in many cases simply wrong. Both papers do not consider the possibility of stars having \Teff\ above 20~kK (which is the case for most OB stars according to the traditional definition) with \citet{Andretal18} erroneously assigning low values for \AGG\ for OB stars with high extinction and \citet{Andeetal19} yielding erroneous relationships between \GBP$-$\GRP\ and \AGG\ for OB stars in general. Therefore, we recommend their \AGG\ values are not used to correct the extinction of OB stars.

\subsection{Comparing distance estimates}
\label{distanceestimates}

$\,\!$\indent It has been known for a long time \citep{LutzKelk73} that using the inverse of the observed parallax to estimate distances leads to biased or even absurd results (as observed parallaxes can be negative). For the case of OB stars, one of us (JMA) developed a Bayesian formalism \citep{Maiz01a,Maiz05c} with a prior based on a self-consistent vertical distribution of the Galactic disk population measured from Hipparcos parallaxes. The prior consists of a self-gravitating isothermal thin disk and an extended halo population of runaway stars and depends on Galactic latitude but not on Galactic longitude. Here we use it to calculate distances for the final ALS sample with the parameters from \citet{Maizetal08a} and the following details:

\begin{figure*}
\centerline{
\includegraphics[width=0.49\linewidth]{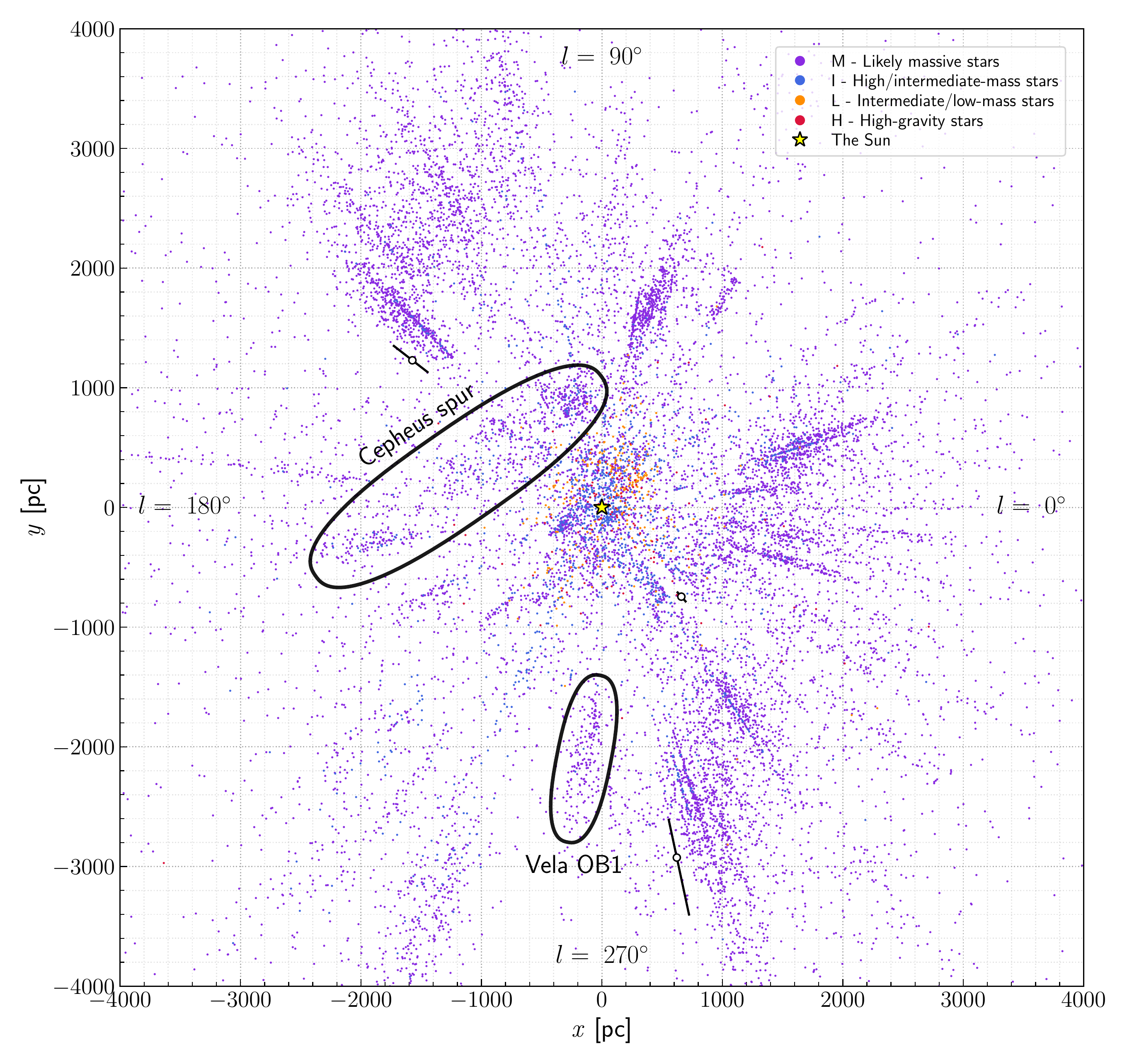} \ 
\includegraphics[width=0.49\linewidth]{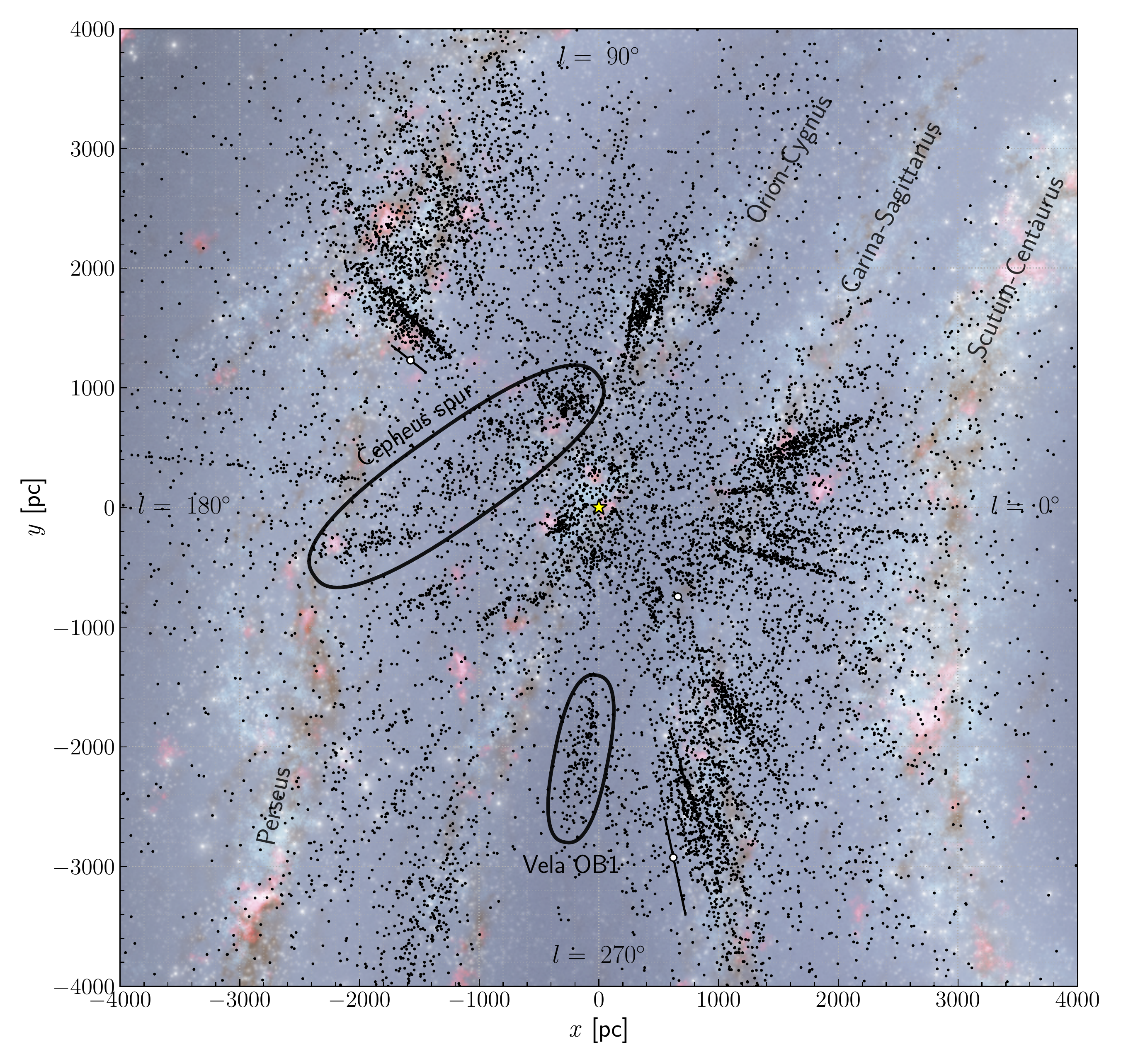} 
}
\caption{(left) Positions of our final sample projected onto the Galactic plane using \dOB\ as the distance and applying the same colour coding by category as in previous figures. Error bars are used to show typical uncertainties for stars located at 1,~2,~and~3~kpc, respectively, and can be used to assess how much of the spread in the radial direction is caused by the uncertainties in \dOB. (right) Version of the left panel with only objects of type M and using the artistic impression of the Milky Way by Robert Hurt as background.}
\label{galmaps}
\end{figure*}

\begin{enumerate}
 \item We apply a parallax zero point of 40~$\mu$as and we add 10~$\mu$as in quadrature to the {\it Gaia}~DR2 uncertainties \citep{Maizetal20b}. 
 \item For objects farther away than 1\degr\ from the Galactic plane, the standard mixture of an isothermal thin disk and a halo (with a fraction of stars of 3.9\%) is used i.e. we consider both possibilities for any given star, that it is not a runaway or that it is.
 \item For objects less than 1\degr\ from the Galactic plane, we use just an isothermal thin disk (i.e. we neglect the possibility of the star being a runaway) and we use the prior corresponding to $b=1\degr$ for stars in the northern Galactic hemisphere and the prior corresponding to $b=-1\degr$ for stars in the southern Galactic hemisphere. This has to be done because the prior was developed using data only from the solar neighborhood and does not consider the possibility that the Galactic disk has a finite extent.
\end{enumerate}

Alternatively, \citet{Bailetal18} and \citet{Andeetal19} use their own Bayesian formalisms to develop more complex priors that depend on both Galactic coordinates. However, those priors are primarily based on the spatial distribution of late-type stars and, in principle, should not be strictly applicable to OB stars, as their distribution in the vertical Galactic direction is narrower than that of late-type stars. In summary, we have three different priors but each one of them has different limitations regarding their applicability to the sample in this paper and, therefore, should yield different distances. We now analyze how large those differences are. 

We want to compare the posterior distributions using the three priors described above. We define \dOB, \dBJ, and \dSH\ as the median of the posterior for each star using \citet{Maizetal08a}, \citet{Bailetal18}, and \citet{Andeetal19}, respectively\footnote{In Table~\ref{cds} we give our distance estimates based on the mean, median, and mode, and \GGabs\ from the median and the mode.}. Given how we have selected the final sample, we can calculate \dOB\ for all of the objects in our final sample except for the seven stars in the Magellanic Clouds, that is, for \num{15656} stars. The same is true for for \dBJ. However, for \dSH\ values can be calculated only for \num{13845} stars. Figure~\ref{distcomp} shows the comparison between \dOB\ and \dBJ\ (left panel) and between \dOB\ and \dSH\ (right panel).

The most important result of the comparisons is that all three distances are very similar. In the range between 0~and~1 kpc the differences between \dOB\ and \dBJ\ grow from 0\% to $\sim$1\% and between \dOB\ and \dSH\ they do it from 0\% and 2\%. The effects are systematic and of opposite sign: \dBJ\ values are larger than \dOB\ ones and these are in turn larger than \dSH\ ones. However, the typical standard deviations of the posterior distributions (in relative terms) are $\sim$2\% around 100~pc and $\sim$5\% around 1000~pc (the plotted values in Fig~\ref{distcomp} are $\sim\sqrt{2}$ larger to consider the contibution from both measurements) i.e. significantly larger. Between 1~and~3 kpc the dispersion of the difference between the compared values increases considerably but so do the standard deviations. In the comparison between \dOB\ and \dBJ\ there is no significant bias (indeed, the trend with distance seen between 0~and~1~kpc is reversed until the bias neutralizes around 3~kpc). In the comparison between \dOB\ and \dSH\ the values for the second become clearly smaller (but within the typical standard deviations of the posterior distribution). Beyond 3~kpc the posterior distributions become very broad but the comparison between \dOB\ and \dBJ\ values is reasonably good. Therefore, our conclusion is that the distances obtained from the OB prior and from the Bailer-Jones prior are very similar and that for the StarHorse distances the differences are somewhat larger but the comparison is still reasonably close. The lesson is that if your prior is a reasonable approximation to reality, your distances will be mostly independent of the details of the prior itself. 

\subsection{Mapping the solar neighbourhood}

$\,\!$\indent In this subsection we use the \dOB\ distances from the previous one to map the location of the OB stars in the solar neighbourhood. Two previous papers have done a similar study cross-matching the previous version of the ALS catalog with {\it Gaia}~DR2. \citet{Xuetal18} applied a straightforward cross-match with a 1\arcsec\ radius and obtained a sample of \num{5772} objects. \citet{Wardetal20} used a larger search radius of 5\arcsec\ that was later cleaned using a $V-\GG$ cut, leaving them with a sample of \num{11844} stars. Those numbers should be compared with our significantly larger sample of \num{15662} objects obtained using the procedure described above.
A third paper, \citet{Zarietal18} analyzes the distribution of young populations in the solar neighbourhood but its scope is different from ours, as it deals mostly with intermediate- and low-mass stars within 500 pc.

\begin{figure*}
\centerline{
\includegraphics[width=0.49\linewidth]{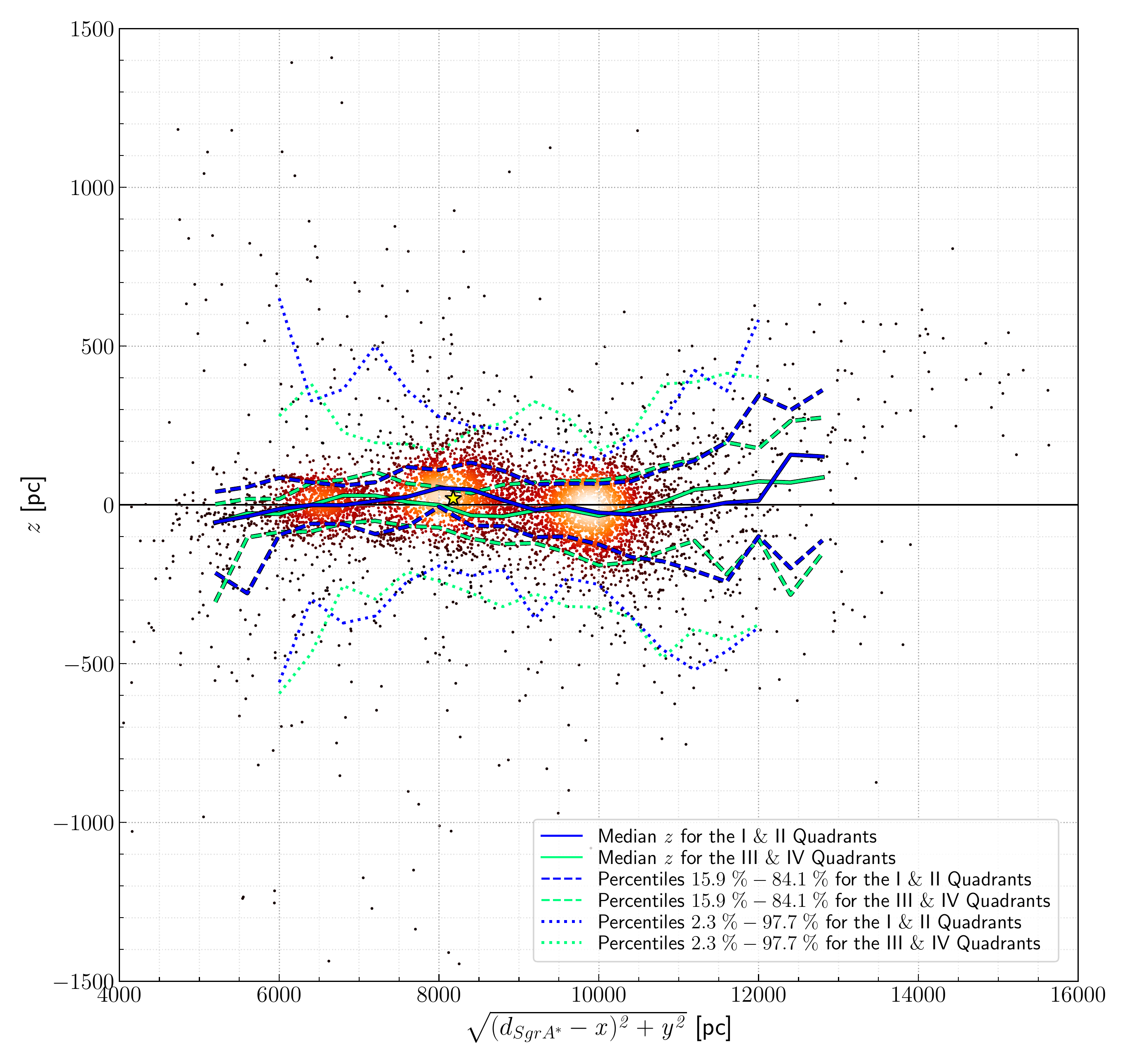} \ 
\includegraphics[width=0.49\linewidth]{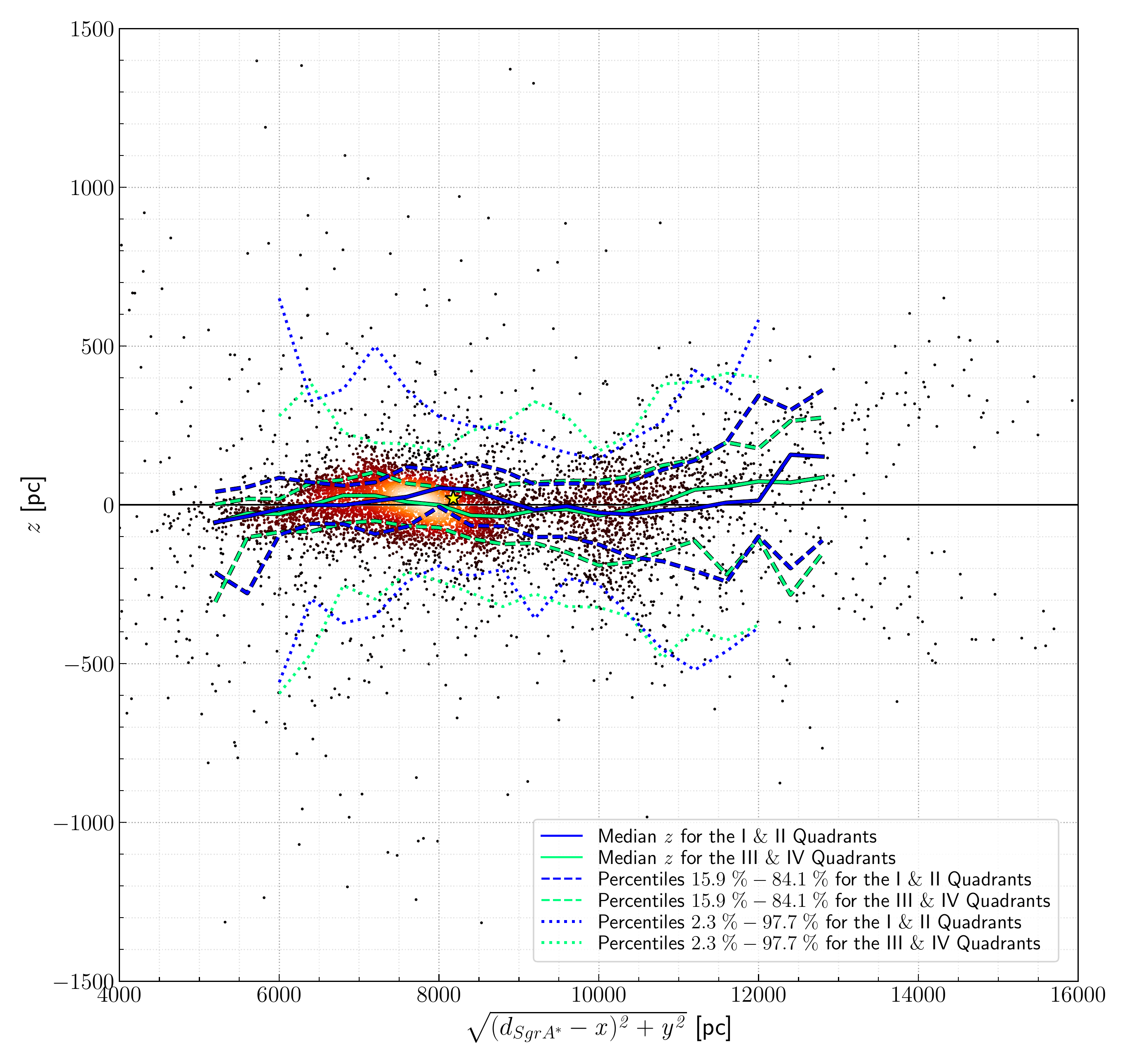} 
}
\caption{Positions of the M-type sample as a function of distance from the Galactic Centre and height with respect to the mid-Galactic Plane (with the Sun located 20~pc above it). The left panel shows the positions for objects in the first two Galactic quadrants and the right panel the equivalent for objects in the last two Galactic quadrants. The running median plus 1-sigma and 2-sigma equivalent percentiles are shown in both plots (blue lines for the distribution in the first two quadrants, green lines for the distribution in the last two). The yellow star marks the Sun's position/}
\label{vertdis1}
\end{figure*}

We first describe the distribution of our sample in the Galactic plane, using as reference Fig.~\ref{galmaps}. In addition, we also provide an animation as supplementary material to better visualize the 3-D structures seen in the data
(see Fig.~\ref{movieframe} for a frame of the animation).
We establish a coordinate system with $x$ and $y$ centered at the Sun's position and where the Galactic Center is at (8.178,0)~kpc \citep{Abutetal19} and where $+y$ is the direction of Galactic rotation. The origin in $z$ is fixed at a position of 20~pc below the Sun's position \citep{Maizetal08a} so that it corresponds to the Galactic mid-plane. The left panel shows the distribution of the stars in our final sample and the right panel only that of M-type objects with the artistic impression of the Milky Way by Robert Hurt in the background (see \url{https://www.eso.org/public/images/eso1339g/}).

The first description of the Galactic spiral structure using OB stars was done by \citet{Morgetal52}. Since then, tracers at different wavelengths have been used to study the configuration of the spiral arms \citep{Vall17,Vall20b}. In our type-M sample we can see delineated three spiral arms. Outside of the solar circle, the Perseus arm is well traced by OB stars in the $l = 100\degr-140\degr$ range but from that point there are few objects \citep{NeguMarc03}. The Orion-Cygnus or local arm is seen in opposite directions of the sky, extending well into the third quadrant between $l=240-250\degr$, in agreement with \citet{Vazqetal08}, and not merging with Perseus arm as previously thought, in agreement with \citet{Xuetal18}. Inside the solar circle, the Carina-Sagittarius arm is the best traced of the three (once one accounts for the artificial spread in the radial direction caused by the uncertainties in \dOB), as expected by the richness of the different star formation episodes present within it (e.g \citealt{Sotaetal14,Maizetal20b}). Beyond the Carina Sagittarius arm, the Scutum-Centaurus arm is not seen in the distribution of OB stars in the ALS catalog mostly due to the strong extinction present in the directions close to the Galactic Centre.
These three spiral arms are the same ones that are seen in the spatial distribution of the nearby OB associations in Fig.~19 of \citet{Wrig20}. Note, however, that Fig.~\ref{galmaps} here reaches a distance twice that of the one shown in that paper.

There are at least two interarm structures seen in Fig.~\ref{galmaps}. One is seen at a distance of $\sim$~2~kpc around $l = 265\degr$ and is produced by the Vela~OB1 association. Its ease of identification is likely favoured by the small amount of dust present in this intearm sightline but note that, as with other structures at those distances from the Sun, the elongation in the radial direction is an artifact of the individual distance uncertainties for each star (see the plotted typical error bars). Note, however, that these structures in Vela have been confused in the past with the continuation of the Orion-Cygnus arm (\citealt{Vazqetal08} and references therein). The second one is a spur that starts from the Cygnus arm at around $l = 90\degr$ and moves out towards the Perseus arm, meeting it around $l = 190\degr$. This structure is delineated by six OB associations: Cep~OB2+OB3+OB4, Cam~OB1, Aur~OB1, and Gem~OB1 and, to our knowledge, it has never been identified before. We name it the Cepheus spur and we discuss it further below. In this case the elongation in the radial direction is smaller because most of its stars are closer than 2~kpc.

We now describe the distribution of OB stars in the vertical direction, which one of us analysed previously using {\it Hipparcos} data \citep{Maiz01a} but without paying attention to the possible changes as a function of position in the Galactic disk. In that respect, the most noticeable structures expected in the Galactic disk are (a) the warp previously detected in positions and proper motions \citep{Reed96,Poggetal18} and (b) possible corrugation patterns, which should be more prominent in young populations than in old ones \citep{MattUson08}. The measured line of nodes of the warp \citep{Poggetal20} is not distant from the anticentre position, with the second quadrant bent upwards and the third quadrant downwards. To check the possible presence of the Galactic warp in our data, we show the vertical distribution as a function of Galactocentric distance in Fig.~\ref{vertdis1}. OB stars in the second quadrant indeed appear to be located preferentially above the plane at large Galactocentric distances and the disk itself appears to become broader as we move outwards in the Galaxy there, in accordance with the predictions of the Galactic warp model of \citet{Poggetal20}. The situation in the third quadrant is less clear, as the vertical distribution seems to bifurcate into an upper and a lower components. It is possible that a small part of the bifurcation is caused by extinction but that is not likely to be all of the story, as dust is scarce in the outer disk. 

\begin{figure}
\centerline{
\includegraphics[width=\linewidth]{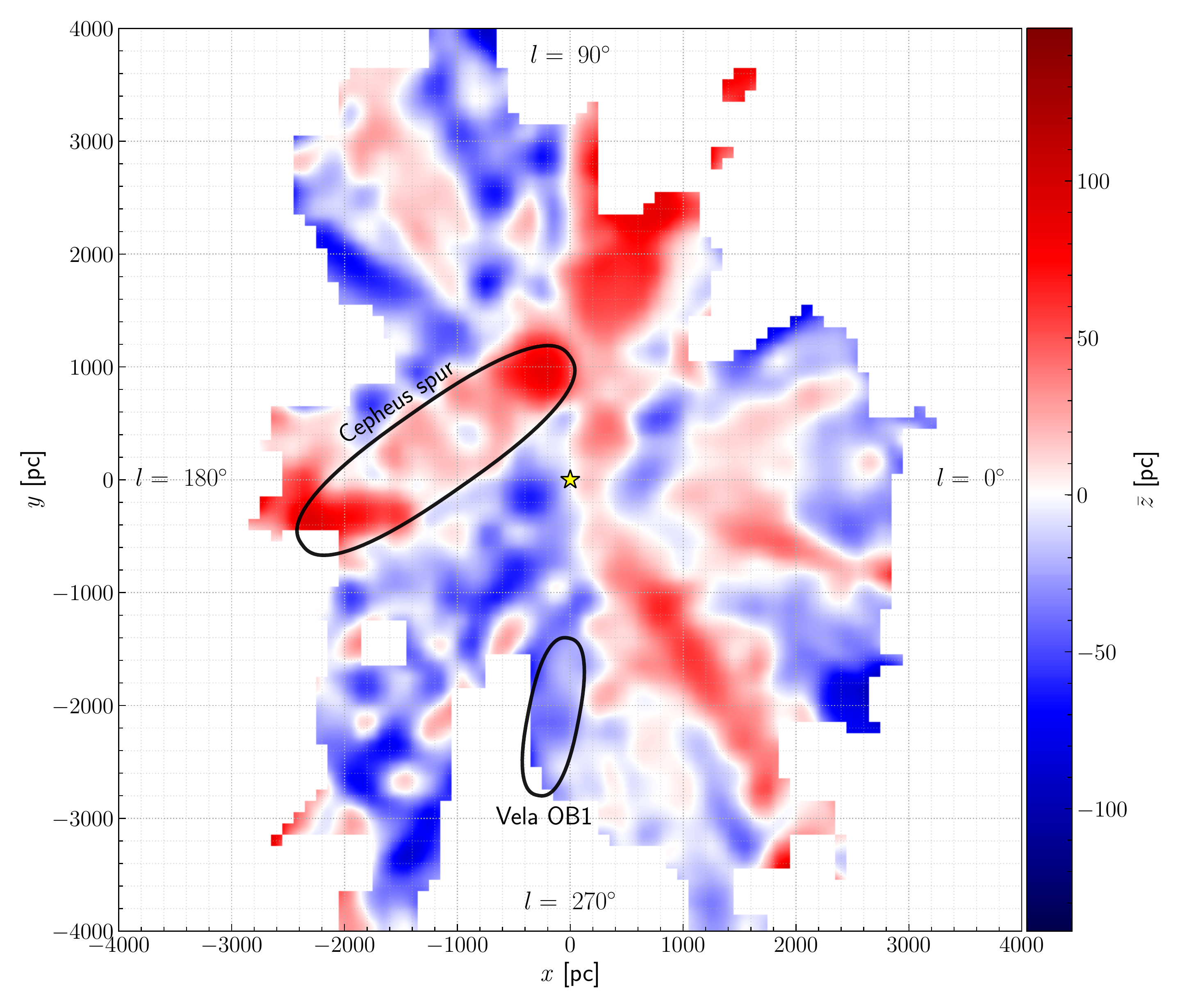} 
}
\caption{Average height with respect to the Galactic mid-plane of the M-type sample as a function of position on the plane. The values have been calculated using a 2-D gaussian smoothing kernel with $\sigma$~=~100~pc. Pixels with a small number of stars have been whitened out to show only regions where the average is calculated using a significant number of stars.}
\label{vertdis2}
\end{figure}

There are other effects seen in Fig.~\ref{vertdis1} that deserve attention. Inside the solar circle the distribution reaches shorter distances that in the opposite direction, a consequence of the rapid increase in extinction towards the inner disk. Also, the distribution shows an increased number at higher latitudes in those directions, indicating there are more runaways in the first and fourth quadrants. This is a likely consequence of the overall higher number of OB stars within the solar circle: extinction may not let us see the OB stars close to the Galactic plane but can do little to avoid the detection of the ejected stars far from it. Another effect in Fig.~\ref{vertdis1} is the different radial distribution between the two panels. In the right one the population is dominated by the Carina-Sagittarius arm while in the left panel the three main arms discussed in this paper (Carina-Sagittarius, Orion-Cygnus, and Perseus) are seen as distinct concentrations at different average Galactocentric distances. A third effect is related to the recent discovery by \citet{Alveetal20} of a structure they dubbed the Radcliffe wave. It is a damped sinusoidal 1-D vertical oscillation that traces the Orion-Cygnus arm, with a region with positive $z$ close to the Sun being above the mid-Galactic plane and a region with negative $z$ in the opposite direction (that includes Ori~OB1) below the mid-Galactic plane. \citet{Flec20} has proposed that such a wave may arise as a result of a Kelvin-Helmholtz instability between the Galactic disk and the Galactic halo. The effect of the Radcliffe wave can be seen in the different behaviour of the median height close to the Sun in the two panels: above average in the left panel and below average in the right panel\footnote{This is aided by the fact that the ALS sample is more complete within 1~kpc of the Sun than at longer distances.}.

\begin{figure*}
\centerline{
\includegraphics[width=0.99\linewidth, trim=0.5cm 0.5cm 0.0cm 0.5cm]{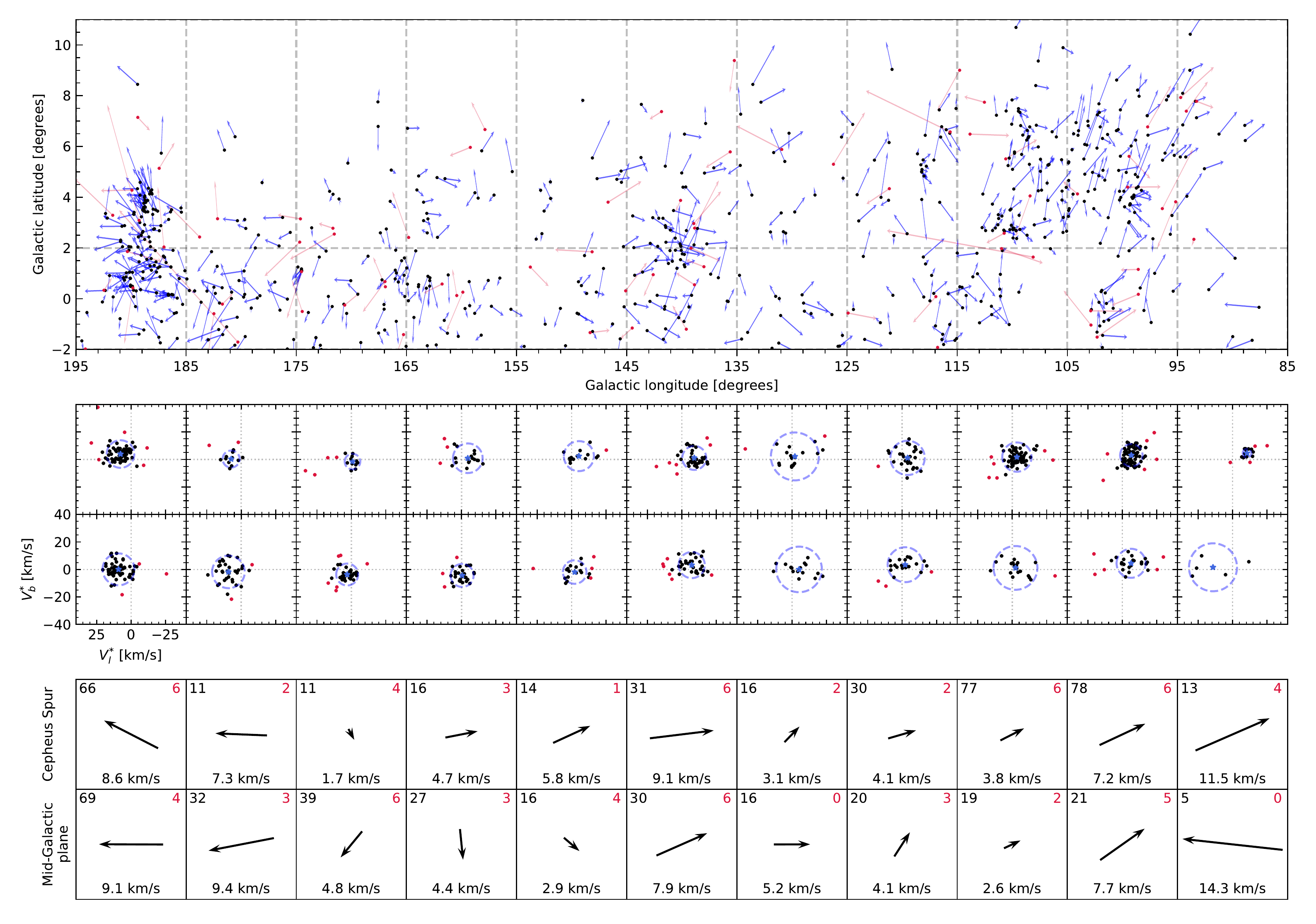}
}
\caption{Kinematic analysis of the Cepheus spur. (top) A sky chart of the M and I catalog stars inside the Cepheus spur oval selection of Fig.~\ref{galmaps}, with arrows showing the peculiar velocities in the plane of the sky. To calculate the velocities we first corrected for the peculiar velocity of the Sun with respect to the Local Standard of Rest and then for a flat Galactic rotation curve model. The red dots and arrows are considered kinematical outliers (possible runaways) and are excluded from the analysis. The grey dashed lines show the limits of the bins used to separate the mid-Galactic plane ($|b| < 2^\circ$) and the Cepheus spur ($b > 2^\circ$) above it in 11 longitudinal tiles. (middle) A representation of the longitudinal and latitudinal components of the peculiar velocity in the 22 bin selections defined in the top chart. The red dots are again the cases excluded during the robust mean calculation while the black dots are used to give an average motion for each bin, which is represented with a blue star. The blue dashed circles correspond to 3 standard deviations from the selected sample of each bin. (bottom) A representation of the robust average corrected transverse velocities of each bin. For each cell the number of stars used for the final average is shown in black, and the number of outliers excluded from the analysis in red. In the lower part of each cell the absolute value of the robust average corrected transverse velocity is given.}
\label{cepspurkinematics}
\end{figure*}

To study in more detail the variations of the distribution of OB stars in the vertical direction, we show the average value of the height with respect to the Galactic mid-plane in Fig.~\ref{vertdis2}. Here we see that the distribution in the vertical direction within a few kpc has a complex behaviour along the $x-y$ plane and does not appear to be dominated by the warp, which can be instead clearly seen for OB stars farther from the Sun (Fig.~\ref{vertdis1}). Instead, the pattern we see is more consistent with being dominated by corrugation effects. A different behaviour in the vertical direction between samples of ``OB'' and RGB stars is also observed by \citet{Romeetal19}. The left panel of their Fig.~5 is the equivalent to our Fig.~\ref{vertdis2} but note that their ``OB'' sample is quite different from ours: much larger in size but including objects with ages up to 1~Ga, while ours contains only ages up to 30~Ma. In other words, most of their sample is made out of intermediate-mass late-B dwarfs, which are not included in the traditional definition of OB stars: they are what we classify as I-type objects.

There are other structures seen in Fig.~\ref{vertdis2} but their reality is unclear at this point without further data. For example, the Cygnus-X region of the Orion-Cygnus arm is also located above the mid-plane but the strong extinction present there questions whether that is just a partial obscuration effect. Two other elevations above the mid-plane in the fourth quadrant are suspiciously elongated in the radial direction (something that does not happen for the Cepheus spur), indicating a possible origin in obscuring clouds on the Carina-Sagittarius arm blocking stars at negative values of $z$ behind them. On the other hand, the already commented Vela~OB1 association is likely to be indeed located below the mid-plane, as in that direction there is little extinction. We note that the structures we see in our sample are quite similar to those seen in Fig.~1 of \cite{Alfaetal91}, who did an analysis similar to ours but using young open clusters. We plan to analyse these corrugation patterns in more detail with {\it Gaia}~EDR3 data and future versions of the ALS catalog with more OB stars.  

\subsection{The Cepheus spur}

\begin{figure*}
\centerline{
\includegraphics[width=\linewidth]{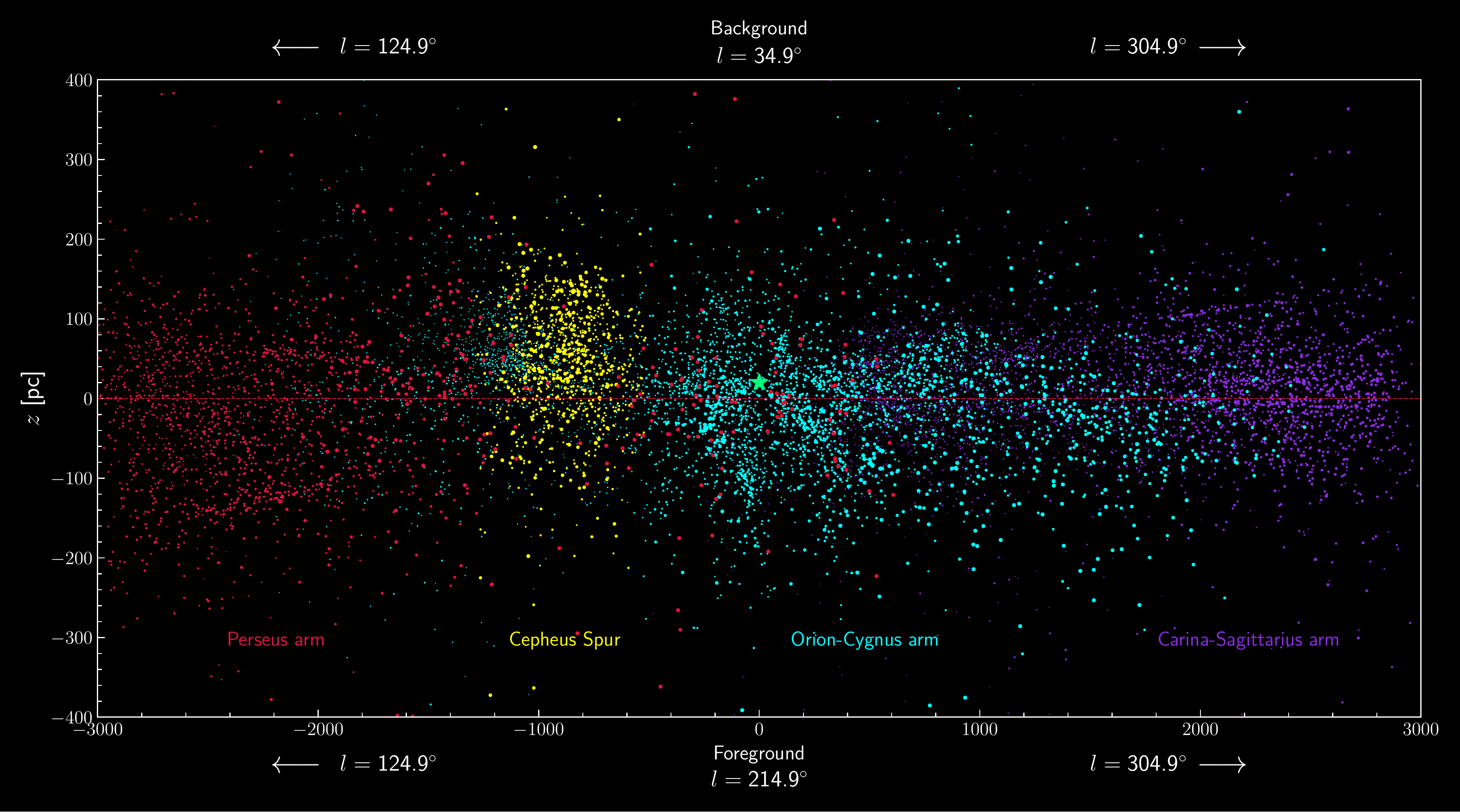}
}
\caption{One of the frames of the supplementary animation associated with this article, where a full rotation around the solar mid-Galactic plane projection point is completed with the Galactic disk shown edge-on. The camera position is at infinity (located in the $l = 214.9^{\rm o}$ direction and pointing towards $l = 34.9^{\rm o}$) so that the $y$ axis  represents the height above the mid-Galactic plane without field-of-view projection distortions. The stars from the M and I catalogs closer than 3~kpc are depicted for the Perseus arm (crimson), the Orion-Cygnus arm (cyan), the Carina-Sagittarius arm (violet) and the Cepheus spur region (yellow, note that we do not distinguish between objects above and below $b=2^{\rm o}$ as we do in Fig.~\ref{cepspurkinematics} ). The position of the Sun is represented by a green star ($20$~pc above the mid-Galactic plane) and the position of the Galactic center is marked in the animation by a red star. The $y$ axis has been exaggerated by a factor of $3.6$ to better illustrate the distributions of heights in the Galactic disk, with the Cepheus Spur oval selection showing the anomalous average height of this structure.}
\label{movieframe}
\end{figure*}

 $\,\!$\indent The Cepheus spur is a conspicuous structure in Fig.~\ref{vertdis2} (see also Fig.~\ref{movieframe}). It starts around (0,1~kpc), $\sim$500~pc beyond where the oscillation of the Radcliffe wave begins, and maintains an elevated position 50-100~pc above the plane as it moves diagonally towards the lower left direction (the real height of the structure is likely to be closer to 100~pc than to 50~pc, as we have not excluded objects that lie below it). Therefore, it looks like the Radcliffe wave is not a 1-D oscillation that creates a 2-D structure in a vertical plane but instead a full 3-D structure extended towards larger Galactocentric radii, with the Cepheus spur being a region of recent enhanced star formation at the crest of the wave. Considering that Ori~OB1 is located at the opposite side of the wave with respect to the Galactic mid-plane (it is the blue region just to the lower left of the Sun in Fig.~\ref{vertdis2}), we tentatively propose that this type of oscillation in the Galactic disk may be responsible for the enhanced star formation at the crests and troughs of the wave. Note that the initial rise of the Cepheus spur between Cygnus and Cepheus was noticed as early as by \citet{Hubb34}, who called attention to the existence of molecular clouds some distance away from the plane and dubbed that part of the structure the Cepheus flare. The associated molecular gas extends well above the heights above the plane where OB stars are found \citep{Kunetal08}. 

To verify the identity of the Cepheus spur we have done an analysis of the peculiar velocities in the plane of the sky for the OB stars in that region of the Milky Way derived from the {\it Gaia}~DR2 proper motions, which is shown in Fig.~\ref{cepspurkinematics}. We selected the sample from the oval region in Fig.~\ref{galmaps} and divided it into an 11$\times$2 grid in Galactic longitude and latitude. The top row corresponds approximately to the Cepheus spur itself while the bottom row is the normal mid-Galactic plane population (we do not show the $b=-2^{\rm o}$ to $-11^{\rm o}$ range because there are few stars in there). As a reference, the dividing $b=2^{\rm o}$ line corresponds to 35~pc at a distance of 1~kpc (right part of the plot) and to 70~pc at a distance of 2~kpc (left part of the plot).

We start the analysis of the Cepheus spur with the peculiar latitudinal motion which, for practical purposes, is very similar to the peculiar motion in the vertical Galactic direction given the small angular distances from the Galactic plane involved. The average value for the stars above $b=2^{\rm o}$ is 2.3~km/s while that for the stars below that value is $-$0.2 km/s. This indicates that the mid-Galactic plane population has the expected behaviour of a near-zero vertical motion while the above-the-plane population is differentiated by having an average upwards motion.

We now turn to the motion in the longitudinal direction. What we see there for $b > 2^{\rm o}$ is the signature of a radiant point similar to that of a meteor shower i.e. the projection of a 3-D velocity vector into a spherical coordinate system. Values are small in the central region of the upper panels of Fig.~\ref{cepspurkinematics} (with the exception of the $l = 135^{\rm o} - 145^{\rm o}$ bin) and increase in opposite directions as we move towards lower and higher Galactic longitudes. If we assume a radiant point at $l = 150^{\rm o}$ and a peculiar velocity in the Galactic plane of 12~km/s toward us, we would see a peculiar longitudinal velocity of $+$7.7~km/s at $l = 190^{\rm o}$ and of $-$10.4~km/s at $l = 90^{\rm o}$, which are nearly identical to the measured values in the two extreme bins of $+$7.7~km/s and $-$10.5~km/s, respectively. For the stars with $b < 2^{\rm o}$ a similar pattern in longitude is seen, so it is possible that the Cepheus spur is a coherent structure that is on average above the plane but extends to a larger range of Galactic latitudes and is characterized by approaching us with a peculiar velocity close to 12~km/s and from a Galactic longitude of $\sim$150$^{\rm o}$. A corollary of this result is that peculiar velocities in the radial direction in the Cepheus spur are expected to be negative. We plan to test that in the future and we also plan to analyze the motion derived from the stellar clusters in the region.

Finally, we analyze the spatial distribution of stars above and below the $b=2^{\rm o}$ dividing line. For $l < 145^{\rm o}$ there are 245 final stars (after excluding possible runaways) above the line and 111 below, a difference of more than a factor of two that is the main reason for the prominence of the Cepheus spur in Figs.~\ref{vertdis2}~and~\ref{movieframe}. In the $l = 145^{\rm o} - 185^{\rm o}$ range there are relatively few stars but the number rises again in the last bin ($l = 185^{\rm o} - 195^{\rm o}$) due to the presence of Gem~OB1. Therefore, we conclude that there are three lines of evidence that support the existence of the Cepheus spur: an overdensity in the $xy$ map, an anomalous average height over the Galactic plane, and kinematics consistent with an overall common peculiar motion.

\section{Future work}

$\,\!$ \indent In the immediate future we plan to incorporate the information from future {\it Gaia} releases. EDR3 will improve the quality of the parallaxes and of the \GBP+\GG+\GRP\ photometry, allowing us to add existing ALS stars to the clean sample in this paper. In DR3 spectrophotometry will become available and with it a better discrimination between source types and a measurement of extinction properties. Also, for some sources (primarily the cooler spectral subtypes) radial velocities will be provided. 

We also plan to add accurate spectral classifications from spectroscopic surveys. For bright stars we have already obtained $R\sim 2500$ blue-violet spectroscopy with GOSSS \citep{Maizetal11} for several thousands of stars. Some of them have already been published (most of them in \citealt{Sotaetal11a,Sotaetal14,Maizetal16}) and can be retrieved from the Galactic O Star Catalog \citep{Maizetal04b} web site: \url{https://gosc.cab.inta-csic.es}. Most of the rest corresponds to B stars and their spectral classifications will be added to future versions of the ALS catalog. We will also incorporate high-resolution spectroscopic results from \lili\ \citep{Maizetal19a}, including radial velocities derived from multi-epoch data. For faint stars spectral types for many OB stars will become available with WEAVE \citep{Dalt16} in the northern hemisphere and with 4MOST \citep{deJoetal19} in the southern hemisphere.

A third type of addition to the ALS catalog will come from ground-based photometric surveys, which can contribute with the $u$-band photometry necessary to identify and characterize OB stars. For this purpose we will use GALANTE \citep{LorGetal19,LorGetal20}, IGAPS \citep{Mongetal20}, and VPHAS+ \citep{Drewetal14}. We may also add {\it Gaia}-identified members from the Villafranca catalog of Galactic OB groups \citep{Maizetal20b} that are bright enough to be OB stars.

The most direct effect of those contributions to the ALS catalog will be a larger sample with many new objects and a more precise characterization of their properties. That in turn will produce an improved knowledge of the spatial distribution of the OB stars in the solar neighborhood, allowing us to detect finer structures and new runaway stars
and to extend the range at which we can discover and study new stellar clusters and associations with OB stars.

\section*{Acknowledgements}

$\,\!$\indent We thank the referee for a constructive report that helped us to improve the paper.
M.P.G. and J.M.A. acknowledge support from the Spanish Government Ministerio de Ciencia through grant PGC2018-\num{095049}-B-C22. 
R.H.B. acknowledges support from DIDULS Project \num{18143} and the ESAC Faculty Visitor Program.
This work has made use of data from the European Space Agency (ESA) Gaia mission, processed by the Gaia Data Processing and Analysis Consortium (DPAC). Funding for the DPAC has been provided by national institutions, in particular the institutions participating in the Gaia Multilateral Agreement. This research has made use of the SIMBAD database, Aladin Sky Atlas and the VizieR catalogue access tool, operated and developed by the Centre de Données astronomiques (CDS), Strasbourg Observatory, France, as well as TOPCAT, an interactive graphical viewer and editor for tabular data, and Astropy, a community-developed core Python package for Astronomy.

\section*{Data availability} 

$\,\!$\indent The data from this paper are available from the CDS.



\bibliographystyle{mnras}
\bibliography{general} 

\appendix

\section{Digging into the previous ALS catalog and cross-matching with Gaia DR2}

$\,\!$\indent The original ALS catalog was the result of years of painstaking data gathering and it provided the largest carefully built catalog of its kind. It is no surprise that there are errors and other issues in the one-by-one elaboration of its \num{18693} entries. Here we acknowledge the diversity of the problems encountered by addressing different types.

{\it Crowding:} NGC~3603 is the richest very young stellar cluster accessible in the optical \citep{Maizetal20b} and, as a result, a severe example of crowding. 
There are 13 ALS sources in NGC~3603 and, except for ALS~\num{2275}, all of those cases lack photometry in the ALS. Since the coordinates were unreliable to perform certain matches, the coordinates shown in the original ALS catalog were disposed in a grid-like fashion around the center of NGC 3603. The cluster also includes two pairs of stars (ALS~\num{19311} / ALS~\num{19314} and ALS~\num{19310} / ALS~\num{19312}) that, in the original catalog, share the exact same coordinates and are not duplicates of each other. With the photographic plates referenced in the ALS, we were able to match 4 of these 13 entries with Gaia DR2 (which could not detect a number of the sources recognizable in this field due to crowding, see \citealt{Maizetal20b}). ALS~\num{2275} (= HD~\num{97950}) in the LS-South catalog refers to the cluster core and its multiple components \citep{Moffetal94}. Some variation of these problems are common in other dense clusters with ALS sources.

{\it Duplicates:} We have discovered 95 instances of duplicates that were not recognised as so in the original ALS catalog, which was expected due to the large overlapping of some of the original references it was based on. In addition there are pairs of stars which are said to be duplicates but are not, like ALS~\num{11110} and ALS~\num{11108}, which are in fact the two components of the binary system HDE~\num{228827}.

{\it Potentially misleading data:}
In the ALS, some values for the $V$ magnitude were intentionally taken from other photometric bands, to better compare the data with the LS photographic plates. Some confusion may arise for ALS~\num{19610} since the $V$-band photometry was instead taken from the column corresponding to the $B$~band in \citet{Chinetal80}. The same happened between the $V$ and $B$ bands for 10 stars in Westerlund 1, taken from \citet{Claretal05a}.

{\it Transcription errors:} Other cases arise from badly transcribed data, either from the reference sources into the ALS or from previous sources into the C-S and LS catalogs. For example, the star ALS~\num{16894} was mistakenly added to the ALS as CD~$-$28~2561 due to a missing letter P in the originally targeted star, CPD~$-$28~2561, which was correctly included as ALS~\num{870}\footnote{As it turned out, this object is a peculiar star, one of the few O stars with a significant magnetic field as signaled by being of Of?p type \citep{Walbetal10a,Wadeetal15}.}. In ALS~\num{19483}, ALS~\num{16986}, ALS~\num{16991}, and many others the $V$ photometry presented in the references was simply ignored. In ALS \num{9528} the coordinates from the LS catalog erroneously substitute the 19 hour in RA with an 18. Similarly, for ALS~\num{17479} the hour in the right ascension was swapped from 17 to 12 and in ALS~\num{19668} the 38 arcminutes in declination were transcribed as 28 by mistake.

{\it Bad-quality coordinates:} Many ALS sources have coordinate uncertainties of a few arcseconds but some can be significantly large. For example, for ALS~\num{12636} (one of the few cases without a Simbad entry) the coordinates are suspiciously rounded to the arcminute. As it turns out, this is possibly a duplicate of ALS~\num{12639} or at least the result of a chain of badly transcribed coordinates from the LS and BD catalogs, which also show large errors in their coordinates.

{\it Simbad:} In some cases Simbad was the wrongdoer and therefore some reexamination of its reliability for the Gaia DR2 cross-match was needed. For example, ALS~\num{15862} was wrongly matched with Gaia~DR2~\num{5350363910256783488}, which in turn is a better astrophotometric match for another ALS source, ALS~\num{1820}, while the best match for ALS~\num{15862} is Gaia~DR2~\num{5350363875897024256}. Simbad also matched ALS~\num{19613} with Tyc~6265-1255-1, which in turn is matched with Gaia~DR2~\num{4097815382164899840} by the external cross-matches in the Gaia archive, an identifier which, according to Simbad, is itself matched to ALS~\num{19618}. Some of these chains of inconsistent cross-matches between ALS references, Simbad, and Gaia DR2 cross-matches with external catalogs are found across the ALS. We have solved these issues adapting to what seemed to be the most plausible scenario in a case-by-case procedure, but in general we have assumed that Gaia DR2 cross-matches have better quality and consistency than those performed by Simbad.

{\it Detective cases:} Finally, in some cases we had to spend a significant amount of time to decipher what was going on. They usually have a combination of issues. Here are some examples.

In just 20\arcsec\ around ALS~\num{18476} there are 30 back-propagated Gaia DR2 sources, many of which are also good photometric matches for ALS\num{18476}. The ALS refers here to some photographic plates published by \citet{Wram80}, where the true match can be visually determined to be Gaia~DR2~\num{5356258185934696576}. However, that paper gives a set of coordinates that are inconsistent with what is shown in the photographic plates, with a separation of 3.7\arcmin\ between them. On top of that, the ALS coordinates differ by as much as 4\arcmin\ from both the true source and the coordinates of its reference at the same epoch.

ALS~\num{19457} was entered into the ALS catalog as star 42 of \citet{OrsaMuzz80}. However, this star is 1.23\degr\ away from the coordinates shown in the photographic plate, from which we derived the true match with Gaia~DR2~\num{5877156797428541952}. The mistake comes from the reference using the declination of the star 43 for the star 42 and the declination of the star 42 for the star 41, while maintaining the correct right ascensions, thus wrongly tabulating the positions of their own photographic plates by displacing this column one entry. The ALS catalog inherited this problem but also ignored the values for the photometry that are correctly shown in the reference, thus making the cross-match analysis even more subtle.

ALS~\num{20122} is designated as NGC~4755~302 in the ALS catalog but there is no 302 star in the reference provided, \citep{Evanetal05}. The true source is in fact NGC~4755~093 in that reference since the classification and photometry coincide with what was really shown in the ALS, but still the coordinates are 7.4\arcmin\ off track here. So, not only the alternate identifier was wrong but also the coordinates were not correctly transcribed from the reference. But, just by changing the minute in the right ascension from 53 to 54, both ALS~\num{20122} and NGC~4755~093 end up in the same place.

Several mistakes in \citet{Wram76b} propagated into the previous version of the ALS catalog. For example, ALS~\num{18375}, which is the star number 3 in the reference, can be recognized as Gaia~DR2~\num{2006098517247905152} in the photographic plates while in the their own table it is located 53\arcsec\ away from that position. This is due to a misalignment between the photographic plate and the superimposed grid of coordinates, which apparently was displaced almost 1\arcmin\ to the southwest direction. This is easily recognizable for ALS~\num{18375}, but when we consider ALS~\num{12201} it results in a problematic situation. ALS~\num{12201} is star number 5 in the reference, but its position in the photographic plates is better matched with star number 3 because of the aforementioned misalignment (when in reality we are dealing with a different star). To make things even worse, star 5 and 3 are very similar in terms of brightness, which could lead to even more confusion about their identity. Because of these issues, another entry, ALS~\num{18832}, which was linked to the star number 162 in \citet{Barbetal73} by the ALS, was not flagged as a duplicate, while in reality it can be seen in the photographic plates that it is the same star as number 3 of \citet{Wram76b}.

Ambiguous markings in photographic plates can also lead to wrong identifications, like in the case of ALS~\num{18099}, also known as star number 77 in \citet{Mart72}. The ALS coordinates would make a good match for Gaia~DR2~\num{2012999082277360512}, but the real position of star 77 in the photographic plates of the reference is confusing due to the printed arrow pointing to this and to another star in a very crowded image. We believe that the real match here is Gaia~DR2~\num{2012993554649161344} which is the same star as ALS~\num{13252}, and thus ALS~\num{18099} can be now recognised as its duplicate. Another problematic star is ALS~\num{18665}, also known as S-237 4 in \citet{Moffetal79}, because the photographic plate associated with it displays a configuration of stars that we could not confidently recognise inside the bright nebula.

\section{Testing the two Gaia DR2 \GBP\ passbands}

$\,\!$\indent One of the features of the {\it Gaia}~DR2 calibration by \citet{MaizWeil18} is the use of two different passbands and two different zero points for \GBP\ depending on the value of \GG\ (not on the value of \GBP) for the target. The need for this is explained by Fig.~4 in that paper, where it is shown that a jump in the sample standard deviations for the \GBP\ flux happens at \GG~=~10.87~mag and that the size of the jump depends on the colour of the source: it is large for blue objects and small for red ones. The clean ALS sample obtained in this paper allows us to test this effect by removing the (vast) majority of intrinsically red stars in the full {\it Gaia}~DR2 sample and leaving only blue stars with different degrees of extinction. Such a selection produces an easy-to-understand sample of intrinsically-blue SEDs that follows (mostly) an extinction sequence.

According to \citet{MaizWeil18}, the differences between the photometry for bright and faint stars can be described by two effects. First, for stars with Vega-like SEDs ({\it Gaia} magnitudes are expressed using Vega as the reference SED) the difference in \GBP\ corresponds to 26~mmag, in the sense that stars fainter than \GG~=~10.87~mag have measured \GBP\ magnitudes that are brighter (for their SEDs) when compared with brighter stars. Putting it in another way, a faint Vega-like star has a {\it Gaia}~DR2 \GBP$-$\GGp\ colour that is 26 mmag bluer than that of its bright equivalent. Second, for faint stars \GBP\ is significantly more sensitive to the left of the Balmer jump than for bright stars. The effect is apparent in Fig.~11 of \citet{MaizWeil18} by comparing the left and right panel. In principle, this could be quantified as a correction using a Johnson $U-B$-like colour (which measures the strength of the Balmer jump) but such a colour is very different to the ones available from current {\it Gaia} photometry. Both \GBP$-$\GGp\ and \GGp$-$\GRP\ are monotonously decreasing functions of \Teff\ but $U-B$ has a more complex behaviour, decreasing for low and high values of \Teff\ but with the opposite behaviour at intermediate values that, furthermore, is a function of gravity. That was the primary reason why \citet{MaizWeil18} had to define two different bands for \GBP.

In order to further test the differences between the two magnitude ranges, we plot in the right panel of Fig.~\ref{colcol} a zoom into the lower right corner of the left panel with the addition of extinction tracks for 40~kK MS stars and \Teff\ sequences for zero-extinction MS stars. The first noticeable effect is that the two magnitude ranges are separated in the colour-colour diagram, with the faint stars to the left, as expected. Also as expected, the difference is larger for stars with bluer colours, as there is where the contribution to \GBP\ from photons to the left of the Balmer jump constitute a larger share of the total. As the SEDs become redder, the two sets become closer together and the large separation in \GBP$-$\GGp\ between the zero extinction sequence for O stars becomes smaller and smaller. 

Another interesting effect in the right panel of Fig.~\ref{colcol} is the separation between the extinction and temperature sequences, which is significantly larger for faint stars than for bright stars. This is a consequence of the larger sensitivity of \GBP\ to the left of the Balmer jump for faint stars. In a sense, for faint stars the \GBP$-$\GGp~+~\GGp$-$\GRP\ DR2 diagram crudely resembles the classical Johnson $U-B$~+~$B-V$ diagram \citep{JohnMorg53}, with separated sequences for extinction and temperature, while for bright stars the similarity is harder to notice. We say ``crudely'' because the separation between sequences is much smaller for the {\it Gaia} diagram than for the Johnson one, as for \GBP\ (the $U$ equivalent) most of of the flux originates to the right, not to the left, of the Balmer jump. If we are able to see these effects is due to the extraordinary photometric quality of the {\it Gaia} data. 

A final aspect of the right panel of Fig.~\ref{colcol} is the location of the high-gravity stars. They belong mostly to the faint sample and they are the heavily dominant contribution to the bluest stars there. This is a consequence of the differences in luminosity between massive OB stars and subdwarfs/white dwarfs. The former are luminous objects located mostly in the Galactic plane. Therefore, if they have blue colours they must be relatively nearby and hence bright. If they are faint, they must have considerable extinction. On the other hand, high-gravity stars have a more uniform distribution in Galactic latitude and can be at the same time relatively nearby and faint.  

\bsp	
\label{lastpage}
\end{document}